\documentclass{aa}  

\usepackage{graphicx}
\usepackage{subcaption}
\usepackage[normalem]{ulem}
\usepackage{txfonts}
\usepackage{siunitx}
\usepackage{multicol}
\usepackage{xcolor}
\usepackage{hhline}
\usepackage{lipsum}
\usepackage{lscape}             
\usepackage{placeins}  
\usepackage{float}
\usepackage[hidelinks,colorlinks=true,linkcolor=blue,citecolor=blue]{hyperref}

\begin{document}

   \title{Accuracy of electric current inference assuming three-dimensional hydrostatic equilibrium}

   \author{Aaron Meissner
          \inst{1}, Adur Pastor Yabar \inst{1}, Jaime de la Cruz Rodr\'iguez \inst{1}
          }

   \institute {Institute for Solar Physics, Dept. of Astronomy, Stockholm University, AlbaNova University Centre, SE-10691 Stockholm, Sweden\\
              \email{aaron.meissner@astro.su.se} }

\abstract
    {The chromospheric and coronal heating problem is among the major open questions in solar physics. There are many possible mechanisms that can explain the rise in temperature that is required to reproduce the observations, many of which are related to the presence of magnetic fields. However, their observational confirmation and relative contributions are far from being fully established.}
    {Studying some of the heating processes in the solar atmosphere requires the estimation of electric currents. Since these depend on the spatial derivatives of the magnetic field components, it is necessary to obtain the full magnetic field vector from observational data in a 3D geometrical grid.}
    {We developed a new method to obtain the 3D geometrical configuration of the atmosphere using the approximation of hydrostatic equilibrium in three dimensions to determine the gas pressure and density as a function of optical depth, column mass, or geometrical distance. This development has been designed as an extension for any inversion code that outputs a stratification of temperature and gas pressure or density to enable the estimation of electric current densities from observations.}
    {We investigated our approach using a snapshot from a sunspot MHD simulation. The difference between the solver output and the simulation, taken as the ground truth, is mostly between $30$ and $45$ \%. It does show a similar pattern over a large optical depth range. We also demonstrate that the system of equations has a unique solution. Importantly, we show that with this approach, the sensitivity to the boundary condition is very small, by comparing results for test runs with different field-of-view sizes. Finally, we also consider the impact of the viewing angle on the results and show that it has only minimal influence.}
    {We conclude that the electric currents can be inferred at a reasonable accuracy with this approach using the simulation
parameters as input. We also tested set-ups that aim at mimicking more realistic cases, where the code performed equally well. The results in this work show promise with respect to future applications to spectropolarimetric data, however, some additional challenges will need to be adressed when moving from the controlled input of the simulation to observational data.}

   \keywords{The Sun -  
   Sun: chromosphere - 
   Sun: magnetic fields - 
   Methods: Data analysis - 
   Methods: numerical
               }
\titlerunning{Inference of electric currents}
\authorrunning{Meissner A., Pastor Yabar A., de la Cruz Rodriguez J.}
   \maketitle

\section{Introduction}
While the study of the solar atmosphere has seen good progress both in observations using modern telescopes and instruments and in state-of-the-art simulations \citep[see e.g.][for reviews on specific topics]{Leenaarts2020,Carlsson2019,Lagg2017,Pontin2022}, there are a number of open questions remaining. One of the major open questions in solar physics is the chromospheric and coronal heating problem. This refers to the fact that the temperature in the chromosphere of the Sun is higher than what we would be able to explain solely considering radiation transport as the heating mechanism \citep{Athay1976}. Additionally, measurements of the amount of energy that is radiated away by the chromosphere are shown to be higher than what would otherwise be expected. Indeed, original studies quantified the average radiative losses in the chromosphere to be around $5$ \si{\kilo\watt\per\square\metre} in the quiet Sun and around $20$ \si{\kilo\watt\per\square\metre} in active regions \citep[e.g.][]{Withbroe1977,Vernazza1981}. More recent studies, using up-to-date instruments with a much higher resolution, have found local rates of up to $160$ \si{\kilo\watt\per\square\metre} \citep{DiazBaso2021} or temporal variations \citep{2022A&A...664A...8M}, further highlighting the importance of determining the contributions of individual mechanisms and comparing them to the radiative losses.

Over recent decades, a wide range of mechanisms have been proposed, which can be broadly categorised into two groups. The first one contains processes that relate to waves and instabilities \citep[see][for a review]{Srivastava2021}, such as magnetoacoustic waves \citep[e.g.][]{2011ApJ...736....3V,Abbasvand2020}. The second one describes the dissipation of electric currents through processes such as Ohmic heating \citep{Louis2021,SocasNavarro2005,Tritschler2008} and, extending to the generalised Ohm's law, ambipolar diffusion mechanisms \citep[see e.g.][]{Nobregar_Siverio_et_al2020,Khomenko2012,Shelyag2016}. Even though many of them have been found in both observations and simulations to contribute to the heating \citep[see][for some examples]{Abbasvand2020,Morton2012,delaCruzRodriguez2013,Brady2016}, the specific contribution of each mechanism at different features and depths has not been well established thus far.

A popular method for gaining insights into the physical parameters of the solar atmosphere from observations is the inversion of spectropolarimetric data using models stratified along the line of sight. Inversion codes typically minimise a metric that compares a synthetic spectrum produced by a radiative transfer module with the observations \citep[see][for a review]{delToroIniesta2016,delaCruz2017}. This approach makes use of data taken from facilities equipped with spectropolarimeters, which are observing spectral lines that probe the solar atmosphere at different heights. Common examples that are used to constrain the magnetic field vector in the photosphere and chromosphere are the \ion{Fe}{I} line at 6173 Å (hereafter referred to as $\mathrm{\lambda}$ 6173) and the \ion{Ca}{II} line at 8542~\AA\ (hereafter referred to as $\mathrm{\lambda}$ 8542) lines. Some examples of these inversion codes are STiC \citep{delaCruzRodriguez2019b}, SNAPI \citep{Milic_et_al2018}, NICOLE \citep{SocasNavarro2015}, DeSIRe \citep{RuizCobo2022}, and TIC \citep{Li2022}. All these codes operate under the assumption of 1D hydrostatic equilibrium, while also adopting a vertical depth scale given in column mass or optical depth. These two assumptions allow for an individual processing of all pixels in the FOV as horizontal radiative coupling is also ignored. The resulting mapping to a Cartesian vertical axis is inconsistent across the FOV: the origin is unconstrained, while the sampling (set by 1D hydrostatic equilibrium) can also be largely inconsistent. A few exceptions, such as FIRTEZ-dz \citep{PastorYabar2019,borrero2019}, account for horizontal pressure balance and the Lorentz force. These automatically introduce physical coupling between neighbouring pixels and allow for the inversion to be carred out on a fixed (pure) Cartesian grid. 

To estimate the contribution of the aforementioned electric current-related dissipation mechanisms, we need an estimation of the electric currents via $\boldsymbol{j} = \frac{c}{4\pi}\nabla \times \boldsymbol{B}$, which depends on the spatial derivatives. Computing spatial derivatives from a model that has been reconstructed under the assumption of 1D hydrostatic equilibrium is not sufficiently accurate to derive meaningful estimates of the current vector due to the aforementioned limitations of the derived vertical axis \citep[see,][]{2010ApJ...720.1417P}.

In this study, we have developed a module for imposing 3D hydrostatic equilibrium in a Cartesian grid. This additional processing takes as input the stratifications of temperature and gas pressure or density from the spectropolarimetric result and the 3D Cartesian grid from the user. Given these inputs, it determines the 3D gas pressure so that hydrostatic equilibrium is ensured in 3D and the temperature stratification along optical depth is kept constant. We prefer to explore here the potential of this conservative approach because, in contrast to the photosphere, the determination of the full magnetic field vector (and its 180 degree ambiguity resolution) in extended areas of the Sun in the chromosphere is far from being systematic. The method will allow us to calculate the electric currents where observations are providing reliable magnetic field vector inferences, without having to resort to extrapolation methods and approximations. Following this, we can then calculate constraints on the contribution of Ohmic heating and ambipolar diffusion, which additionally requires atomic data such as cross-sections and number densities. Furthermore, it will also enable a comparison between the heating values and the radiative losses in the chromosphere.

We discuss our approach on how to solve hydrostatic equilibrium in 3D and our implementation in Sect.~\ref{sec:2}. In Sect.~\ref{sec:3}, we then show the results of our validation tests using a snapshot from a radiative MHD simulation, which are discussed in Sect.~\ref{sec:4}. Finally, we present our conclusions and provide an outlook on future improvements and plans for the application of this method to real data in Sect.~\ref{sec:5}.

\section{Method} \label{sec:2}

Most radiative magnetohydrodynamic simulations use a resistive equation of motion to resemble the plasma in the solar atmosphere as closely as possible. As a first simplification, the ideal MHD equation of motion can be written as
\begin{equation}
    \rho\frac{\partial\boldsymbol{v}}{\partial t} = \rho \boldsymbol{g} - \nabla P + \frac{1}{c} \boldsymbol{j} \times \boldsymbol{B} - \rho\left(\boldsymbol{v}\cdot\nabla\right)\boldsymbol{v},
\label{eq:MHD}
\end{equation}
ignoring viscosity and ohmic resistance. Here $\rho$ is the density, $\boldsymbol{g}= \left(0,0, -274 \right)$ \si{\metre\per\second\squared} is the solar surface gravity, $\boldsymbol{v}$ is the velocity of the fluid, $\Vec{B}$ is the magnetic field vector, $\boldsymbol{j} = \frac{c}{4\pi}\nabla \times \boldsymbol{B}$ the electric current, and $P$ the pressure. Since we are only able to determine the line-of-sight velocity from the inversions and perpendicular velocities pose strong challenges, it is common to leave these terms out of consideration. Dropping the velocity terms then leaves us with the magnetohydrostatic equilibrium approximation,
\begin{equation}
    \nabla P = \boldsymbol{g}\rho + \frac{1}{c}\boldsymbol{j}\times \boldsymbol{B},
\end{equation}
where gravity, pressure gradients, and the Lorentz force are left in the force-balance. If we further simplify the equation by neglecting the Lorentz force, we arrive at hydrostatic equilibrium in 3D via

\begin{equation}
    \nabla P = \boldsymbol{g}\rho.
    \label{eq:HE}
\end{equation}
Most inversion codes today adopt the vertical component of Eq.~\ref{eq:HE} to calculate a pressure scale, while ignoring horizontal pressure balance between neighbouring pixels. This assumption allows processing pixels from the observation independently, but it also precludes the code from deriving a common vertical scale axis for all pixels.

This is where our method will come into play. It will take the output of an inversion; namely, the temperature and a first guess for the pressure or density. We then use it to solve the first-order system in Eq.~\ref{eq:HE} in 3D. In addition, we need a boundary condition to prevent the atmosphere from globally shifting in the vertical direction. We can use a fourth-order finite difference scheme for the pressure derivatives, turning Eq.~\ref{eq:HE} into a system of linear equations $\mathbf{A}\boldsymbol{P}=\boldsymbol{b}$, using the step size from the input grid. In a system with $N$ number of spatial grid points, $\mathbf{A}$ is a $3N\times N$ matrix containing the coefficients of the finite difference scheme, $\boldsymbol{b}$ is a vector of a size $3N$ containing the gravitational term of the force balance and $\boldsymbol{P}$ the solution with $N$ entries. 

\begin{figure}
    \centering
    \includegraphics[width=\linewidth]{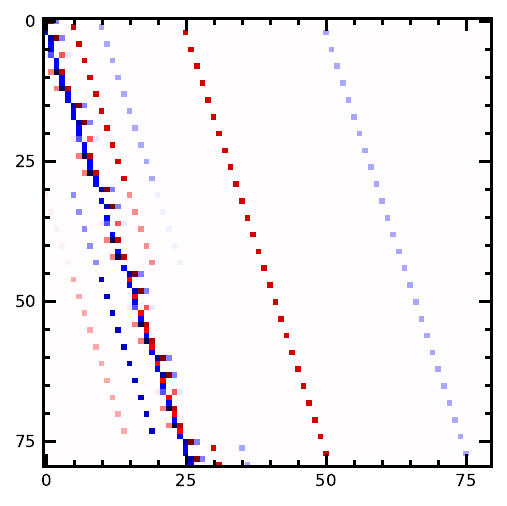}
    \caption{Partial image of the matrix $\mathbf{A}$ and its structure for the case of a $5\times5\times5$ grid. The pixels in blue represent positive values, the pixels in red negative. The matrix was filled according to the fourth-order finite difference scheme, using z,x,y order. We used the third-order forward and backward scheme for the respective boundaries.}
    \label{fig:matrix}
\end{figure}

An example of the $\mathbf{A}$ matrix for a small number of pixels is shown in Fig.~\ref{fig:matrix}, where the derivatives are ordered in z-,x- and y-direction for each pixel. 
The same applies for the right-hand-side vector, $\boldsymbol{b}$, which means that only the entries with the index $0,3,6,...$ are non-zero. Here we also note that we implemented all the equations and operations in logarithm, where we divide $\boldsymbol{b}$ by the pressure from the previous iteration, making use of the logarithmic derivative on the left-hand-side. This was done to both improve stability, as the solver now does not have to handle values spanning multiple orders of magnitude any longer and to ensure the solver will not produce negative values for the pressure. 

Since the system is overdetermined due to the three derivatives per grid-point, we use the set of normal equations to obtain a square matrix, multiplying the system by its transpose. We solve this sparse system of linear equations using iterative methods like the biconjugate gradient stabilized method \citep[BiCGSTAB,][]{vanderHorst_1992}. While the formation of the normal equations squares the condition number, this configuration using the aforementioned solver with the default diagonal preconditioner proved to be the fastest of the implemented solvers in the Eigen library \citep{eigenweb}. We make use of the Eigen library, as it provides an efficient implementation of sparse matrix operations, including a multi-threaded implementation of BiCGSTAB. 

The solution of Eq.~\ref{eq:HE} is a new estimate of gas pressure, whose combination with the temperature and an equation of state (EOS) leads to a new density estimate. Thus, the right-hand-side of Eq.~\ref{eq:HE} is modified and the system has to be solved again, turning into an iterative procedure. Furthermore, this method is designed to be applied after a spectropolarimetric inversion has been performed. If it is carried out as described thus far, the method presented here will change the gas pressure (and density) on the geometrical grid, changing the relation between geometrical distance and optical depth. Consequently, the physical parameters may not have the same optical depth variation as in the input data. Thus, the final atmosphere will not necessarily reproduce the same spectra. To account for this effect, we introduce an additional iterative step within the previous one every time a new gas pressure is calculated. This additional step keeps the optical depth dependence of the temperature identical. In other words, there are two nested iterative procedures: the outer one modifies the gas pressure, while the inner one ensures that the optical depth stratification of the temperature remains identical for every outer loop iteration. Figure~\ref{fig:flowchart} shows a flowchart sketch, summarizing the iterative implementation of the method. Further details of each step are provided below:

\begin{figure}
\centering
\includegraphics[width=\linewidth]{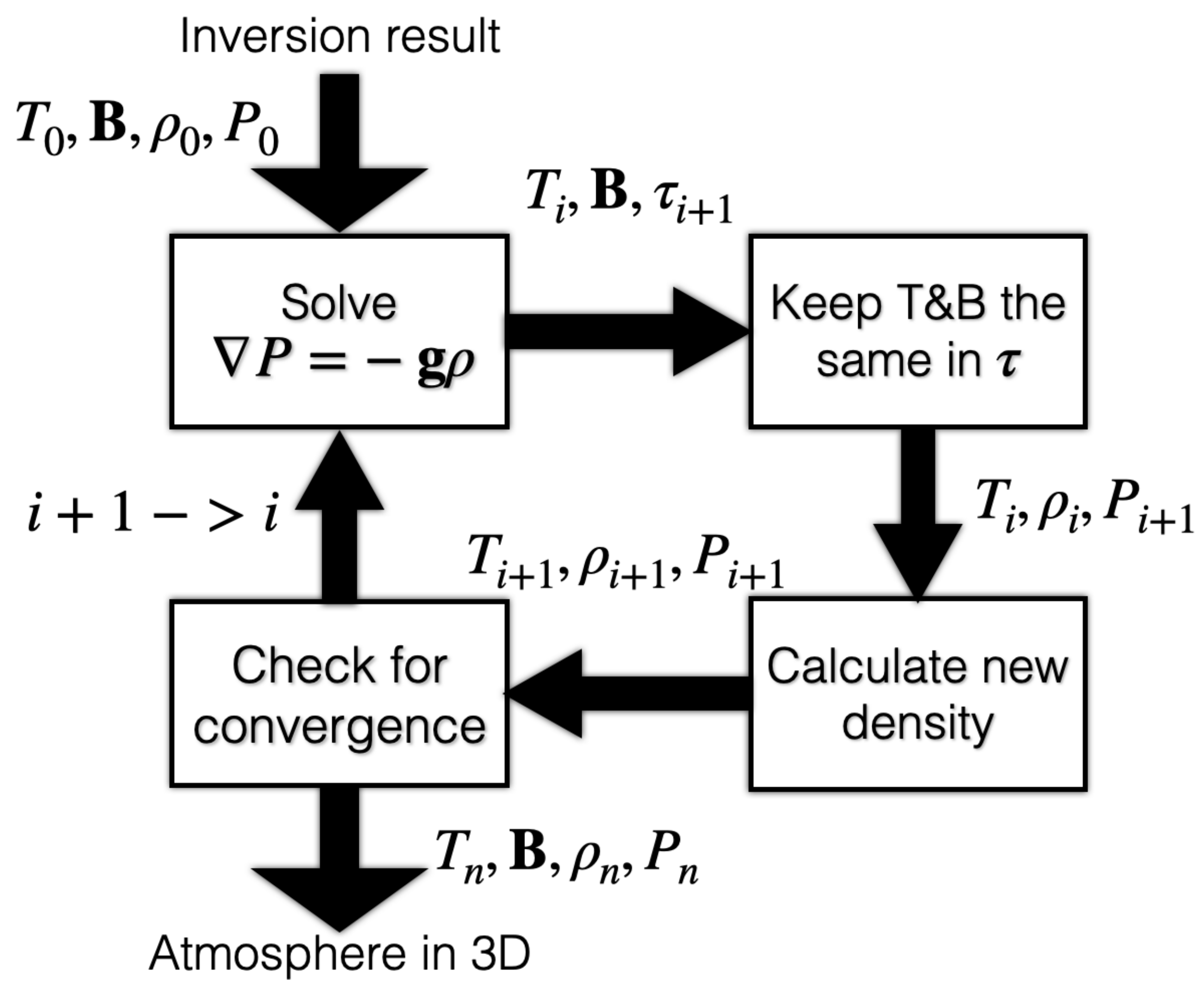}
\caption{Flowchart of the presented module, showing the iterative procedure solving for the pressure and the density. The parameters that are used in each element are also marked with indices that refer to the iteration number.}
\label{fig:flowchart}
\end{figure}

\begin{itemize}
\item Input data: from the output of any spectropolarimetric code, we take the temperature and gas pressure or density, denoted with the index $0$, and a column mass, optical depth, or height scale. These are, strictly speaking, the only mandatory input parameters required to solve HE in 3D. In addition, any additional physical parameter to be considered afterwards has to be provided to keep its optical depth variation (see below).

\item Solve Eq.~\ref{eq:HE}: the solution of the system provides the spatial variation of the gas pressure without an absolute reference; i.e. any constant pressure shift yields an equally valid solution.

\item Set gas pressure reference value: we enforce a pressure reference such that the average pressure of a given optical depth layer is kept constant. The average pressure at this layer is calculated in each iteration. The value is then subtracted from all points of the 3D volume followed by adding the value from the previous iteration, thereby keeping the average pressure the same. The reference optical-depth layer is chosen by the user depending on which lines are used in the input data (i.e. where these lines are sensitive within the model domain). In doing so, we can ensure that the formation region is kept within the volume considered when solving Eq.~\ref{eq:HE}. Specifically, in this work, we chose $\log_{10}\left(\tau_{500}\right)=-0.5$, as we expect that most real applications will include observations that probe the low photosphere.

\item Inner loop: update the geometrical variation of T to keep its optical depth dependence identical. As the gas pressure is iteratively modified, so is the optical depth. To avoid this change in the temperature stratification with optical depth, we re-interpolate the temperature at any iteration $T_{i}\left[\log_{10}\left(\tau_{500}\right)\right]$ to the input one, $T_{0}\left[\log_{10}\left(\tau_{500}\right)\right]$. This is also done via an iterative procedure, where the temperature is interpolated with the new optical depth and then fed into the EOS to update $\tau$ again. This process is repeated until the relative optical depth change is below a given threshold of $0.2\%$. The temperature interpolation will modify the original geometrical stratification by locally stretching or compressing it, but it will keep the optical-depth stratification identical. This step also updates the density to be used in the next iteration if necessary.

\item Check for convergence: this iterative procedure between solving for the pressure and updating the density is done until convergence, which is deemed the case once the change between iterations is below an average relative difference per pixel of $0.2\%$ in our tests, but can be chosen by the user.

\item Remap the remaining inversion parameters: we adjust any additional inversion parameters (magnetic field, LOS velocity, microturbulent velocity) so that they remain the same value at each optical depth point, to ensure that the synthetic spectra produced by the radiative transfer module are the same as they were before solving for hydrostatic equilibrium in 3D.

\item Electric current density calculation: Once converged, the solver essentially yields an updated optical depth based on the stratification of the plasma parameters, which then allows us to compute the spatial distribution of currents using the magnetic field components from the inversion. 

\end{itemize}

\section{Results} \label{sec:3}

We used a snapshot from an r-MHD simulation to evaluate the validity of our approach and investigate the behaviour of the solver. This sunspot simulation was performed by \citet{Rempel_2012} using the radiative MHD code MURaM \citep{Voegler2005}. Our snapshot contains $4096\times512\times256$ data points with a horizontal sampling of $\mathrm{dx,dy} = 12$ \si{\kilo\metre} and a vertical sampling of $\mathrm{dz} = 8$  \si{\kilo\metre}. 

\begin{figure}
\centering
\includegraphics[width=\linewidth]{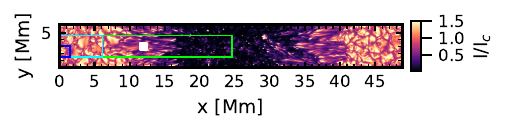}
\caption{Map of surface intensity from simulation snapshot. The green rectangle marks the FOV used for the tests. The blue and cyan colored rectangles mark the sub-fields used when investigating the impact of the chosen size of the data cube. The white square marks the pixel in the penumbra used in the plots that show the variation of current-components with optical depth.}
\label{fig:map_intensity}
\end{figure}

The resulting continuum intensity map is shown in Fig.~\ref{fig:map_intensity}. For our investigations, we limited the data to a field of view (FOV) containing a mix of quiet Sun, penumbra and umbra, with a size of $2048\times256\times256$, highlighted as a lime rectangle in the image. Here, we also marked the two sub-FOVs that were used in one of our tests in cyan and blue, together with a pixel in the penumbra that was chosen to judge the accuracy of electric currents over optical depth, where we expected clear deviations from hydrostatic equilibrium. 

For our assessment, we compared the electric current components from the spatial derivatives of the magnetic field at the same optical depth. As an investigation of the currents is our primary interest, we did not study the plasma parameters with the same level of detail. We note that enforcing horizontal gas-pressure balance is expected to cause the pressure to deviate from that in the simulation, because some terms in Eq.~\ref{eq:MHD} are neglected. Additionally, the horizontal pressure balance requirement means that if the FOV contains regions with very different features (such as the quiet Sun, penumbra, and umbra as in the case presented here), the pressure and density from the solver can differ by several orders of magnitude and should be taken with care. Therefore, we did not consider the gas pressure and density themselves for the comparison and we focussed on the electric currents instead. 

We can use the relative difference to quantify the accuracy, as its value is straightforward to interpret via
\begin{equation}
\frac{|\left(\boldsymbol{j}_{\mathrm{result}}-\boldsymbol{j}_{\mathrm{reference}}\right)|}{|\boldsymbol{j}_{\mathrm{reference}}|}.
\label{eq:symmrelativediff}
\end{equation}
This metric is calculated for every grid point, where we take the median to have a single value characterising the comparison between cases. Since the metric above is not stable where the reference is around zero and the magnetic field determined from observations will contain noise, we masked all the pixels below a certain threshold. We determined the threshold by creating Gaussian noise with a sigma of $10\:(2)$ Gauss for the horizontal (vertical) magnetic field and calculating the electric current components. The numbers we used were chosen as an optimistic lower limit and in real applications of depth-dependent spectropolarimetric inversions, we can expect higher values \citep[see for instance][where they have estimations for the chromospheric magnetic fields of approximately $80\:(70)$ G for 3D spatially regularized inversions and $160\:(140)$ G for more common spectropolarimetric inversions]{arramy2026}. We then used the peak in the distribution of values as the threshold, which was around $0.025 \: \mathrm{M G}\mathrm{s}^{-1}$ for all three components. In the following, this threshold was used over the MHD counterpart to select the currents to be considered in the relative difference calculation. In most comparisons, this meant that we considered roughly $50\%$ of the pixels. Although the chosen noise level is quite optimistic and will, in practice, heavily depend on the chosen model and observational setup, higher noise will further reduce the number of pixels considered in the analysis. In our experience, this will initially lead to a lower relative difference, until a certain electric current value is reached and only the strongest currents are considered; then, the relative difference rises again, as the HE3D prediction usually underestimates the strongest currents (see \ref{subsec:maintest}).

\subsection{Test setup}\label{subsec:listoftest}

In the following, we present the results for the various comparisons that we used to address the accuracy of the method aiming at evaluating the impact of different assumptions. To do so, we considered five different datasets, focussing on the electric current density coming from each of them:

\begin{itemize}
\item MHD case (labelled MHD): in this case, we took the magnetic field vector ($\boldsymbol{B}_{\rm MHD}$), temperature ($T_{\rm MHD}$), and gas pressure ($P_{g,\rm MHD}$) from the r-MHD simulation, denoted with 0 in Fig.~\ref{fig:flowchart}. From these, we computed the optical depth and electric current density ($\log_{10}\tau_{\rm MHD}$ and $\boldsymbol{j}_{\rm MHD}$) compatible with the equation of state, chemical abundance set, and differentiation scheme. We then used these as the ground truth in the comparison with the method described here.

\item Node-sampled MHD case (labelled MHD\_node): in this second case, we followed a similar procedure, but first modifying the $\boldsymbol{B}_{\rm MHD}$ and temperature ($T_{\rm MHD}$) so that they would mimic an observationally driven case in which the inference of these parameters was done using the concept of `nodes' (i.e. a subset of model parameters at a reduced number of grid points along the line of sight from which the model parameters at the radiative transfer grid is interpolated). This case serves as a more realistic scenario for comparison because, in future applications of this method, the initial guess of the parameters will come from observationally driven inferences. To mimic a node-based case, we interpolated the quantities to a grid of equidistant points in height. We chose $13$ points, as this resulted in $4-5$ points in the optical depth range of $1>\log_{10}\left(\tau_{500}\right)>-5$, which is an optimistic, but still realistic value. We then minimized the squared difference between the values at those points and the simulation to obtain a result that is less dependent on the location of the nodes. In this process, the thermodynamics were modified, leading to a new optical depth stratification $\log_{10}\left(\tau_{\mathrm{MHD_{node}}}\right)$; thus, the parameters could then be interpolated back to the original grid. In the process, $\boldsymbol{B}$ was further modified $-$ not only due to the usage of nodes, but also because of the difference in optical depth $-$ thereby leading to $\boldsymbol{j}_{\rm MHD_{node}}$.

\item Solver run (labelled HE3D): in this run, we took the full MHD stratification of the temperature and gas pressure  ($T_{\rm MHD}$ and $P_{g,\rm MHD}$) as input to the solver, as well as the magnetic field vector, $\boldsymbol{B},_{\rm MHD}$ to estimate the electric currents ($\boldsymbol{j}_{\rm HE3D}$). In this case, the HE3D-solver will accept $T_{\rm MHD}$ and $P_{g,\rm MHD}$ as the initial state to work on, denoted with 0 in the flowchart, and end up in a solution characterised by $T_{\rm HE3D}$, $P_{g,\rm HE3D}$, and $\log_{10}\tau_{\rm HE3D}$, denoted as 'n' in Fig.~\ref{fig:flowchart}. 

\item Solver runs on shifted MHD case (labelled HE3D\_shift): the input from the MHD simulation contains horizontal structure, as it employs physical constraints also in the horizontal directions. To remove that horizontal structure in the temperature, pressure, and density as well as test its implications, we applied a shift to every pixel in the input so that $\mathrm{log}_{10}\left(\tau_{500}\right)=0$ is at the same height. In this process, the stratification for each pixel individually in height is preserved.

\item Node-sampled solver run (labelled HE3D\_node): this case is fairly similar to the HE3D case, but instead of taking MHD inputs, we took the node-sampled MHD stratifications from $\mathrm{MHD_{node}}$. The rest of the procedure was kept the same. This case is aimed at checking the limitations the method faces over a more realistic case.

\end{itemize}

\subsection{Accuracy of the solver}\label{subsec:maintest}

\label{sect:he3d_ideal}
\begin{figure}
\centering
\includegraphics[width=\linewidth, trim=0 0.25cm 0 0, clip]{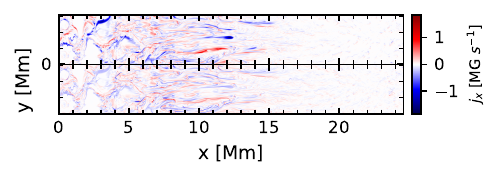}
\caption{Maps of ${j}_x$ at $\mathrm{log}_{10}(\tau_{500})=-2$ for the MHD (upper) and for the HE3D case (lower).}
\label{fig:main_map_HE3D}
\end{figure}

The first and main test was to investigate the accuracy of our method and how much difference the assumption of 3D hydrostatic equilibrium would induce. We use the parameters from the simulation as the input for the solver and compare the resulting output. In Fig.~\ref{fig:main_map_HE3D}, we show maps of the x-component of the currents for the two cases at an optical depth of $\mathrm{log}_{10}(\tau_{500})=-2$, where MHD is on the upper panel and HE3D in the lower panel (maps for the other two components as well as for other optical depth values can be found in Figs.~\ref{fig:AP_main_mapjx-1} to \ref{fig:AP_main_mapjz-3}). It is immediately apparent that most of the spatial features and structures are amply reproduced by the solver output, with the exception of the smallest scales, although the amplitude differs significantly for the larger peaks. To quantitatively assess the performance, we calculated the relative error: for $\mathrm{log}_{10}(\tau_{500})=-2$, the median is $0.31$ (x), $0.38$ (y), and $0.30$ (z). This means that 50\% of pixels have a difference lower than around $30\%$ to $38\%$. The difference is mainly induced by the absence of the additional terms in the equation used in the simulation compared to the employed hydrostatic equilibrium Eq.~\ref{eq:HE} in our solver.

\begin{figure}
    \centering
    \includegraphics[width=\linewidth]{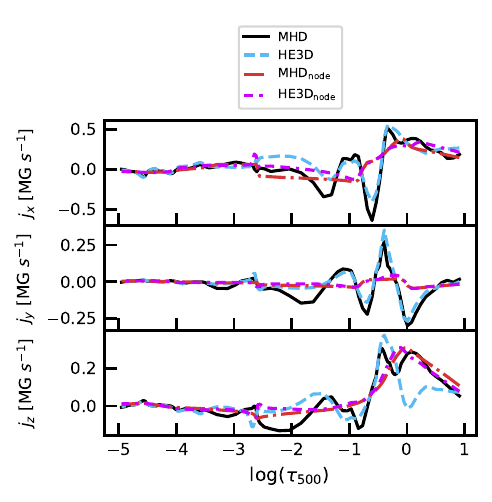}
    \caption{Components of the electric current vector of an individual pixel in the penumbra for an optical depth range of $1>\log_{10}\left(\tau_{500}\right)>-5$, giving the four different cases.}
    \label{fig:main_pix_1000}
\end{figure}

In addition to the comparison of the spatial distribution at specific optical depths, we also investigated the current components in the range of $1>\log_{10}\left(\tau_{500}\right)>-5$ for a pixel in the region of the penumbra (square in Fig.~\ref{fig:map_intensity}). From Fig.~\ref{fig:main_pix_1000}, we see that the solver output, with a few exceptions, follows the same trend as the simulation, but can differ significantly when looking at the exact values. This is expected, as here in the penumbra the other terms contribute significantly to the force-balance, therefore the assumption of hydrostatic equilibrium is least accurate in this part. Similarly to the values at specific optical depths, the median relative difference for this optical depth range are $0.25$ for ${j}_x$, $0.25$ for ${j}_y$ and $0.26$ for ${j}_z$.

\begin{figure}
    \centering
    \includegraphics[width=\linewidth]{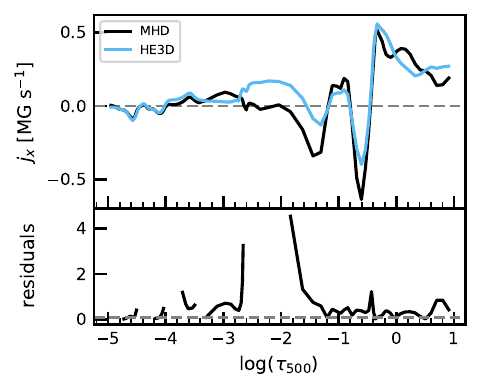}
    \caption{\emph{Top}: ${j}_x$ for an optical depth $1>\log_{10}\left(\tau_{500}\right)>-5$ for both the MHD and HE3D cases. \emph{Bottom}: Residuals between the two curves shown above, where the aforementioned mask is applied, causing the discontinuities.}
    \label{fig:shift_pix_750_reldif}
\end{figure}

Since we know that around $0.01\%$ of pixels show relative differences higher than $100$ even with the mask, we wanted to investigate the current density profiles where these extremely large differences occur. According to Eq.~\ref{eq:symmrelativediff}, these extreme differences must be located where the solver output is strongly overestimating (in absolute value) compared to the simulation. For this purpose, we show the two electric current density (x-component) stratifications for the MHD and HE3D cases together with the relative difference over optical depth in Fig.~\ref{fig:shift_pix_750_reldif}. Here, the residual curve is discontinuous, as we omit the optical depth where MHD current density (in absolute value) is below the threshold. We can see that the highest peaks of residuals are mostly at smaller currents and locations with large gradients of electric current densities. The latter is particularly common, as even a shift of one pixel of the slope location means a very high relative difference even when the MHD value is large. 

To compare the difference between the three components, we also calculated the cumulative distribution, summing up the residuals for all pixels cumulatively, starting from the lowest. Therefore, we know that having smaller differences will introduce a large contribution to the right side of the plot, causing the curve to rise faster. The resulting functions are shown in Fig.~\ref{fig:main_cdf_comp} at $\mathrm{log_{10}}(\tau_{500})=-2$. Other optical depths are given in the appendix, where it becomes clear that there are only slight differences between the three directions, which change between different optical depths. Therefore, there is no conclusive trend seen for any component with respect to being steadily better or worse than the others. 

\begin{figure}
    \centering
    \includegraphics[width=\linewidth]{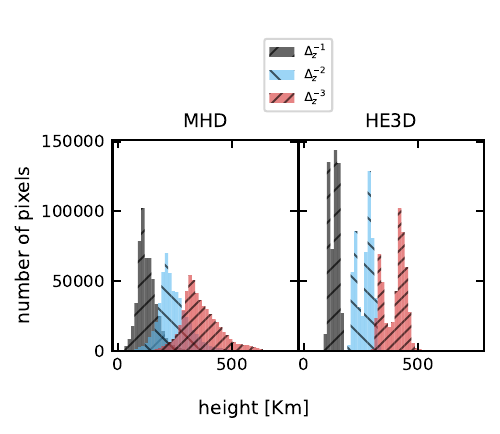}
    \caption{Distribution of height differences between different optical depths for cases of MHD and HE3D.}
    \label{fig:main_hist_z}
\end{figure}

In addition to the accuracy of the currents with respect to the optical depth, we also investigated whether the solver output is able to place the currents at the correct relative height. To do so, we calculated the absolute difference of the height at which a specific optical depth is found compared to the height at which the reference optical depth is located: $\Delta_z^{i} = |z\left[\log_{10}\left(\tau_{500}\right)=i\right] - z\left[\log_{10}\left(\tau_{500}\right) =0\right]|$). Therefore, we looked at the distribution of height differences between different optical depth layers ($i=[-1,-2,-3]$) for the MHD and HE3D cases (Fig.~\ref{fig:main_hist_z}). While the distribution of relative height differences in the MHD simulation is wider than their counterparts for the HE3D case, the majority of height differences are similar. Thus, the electric currents are generally set at the correct heights relative to each other but with a smaller spread in heights. This difference is likely caused by the flattened pressure gradients in the horizontal directions. A striking property of the HE3D height difference distributions is their double-peaked shape. This structure is a consequence of having narrower distributions for three distinct populations associated with the quiet Sun, penumbra, and umbra (as shown in Fig.~\ref{fig:AP_main_hist_z_reg}) that show smaller differences for the umbral regions and larger ones for quiet Sun and penumbral areas. Interestingly, the spatial distribution of the absolute change in height difference between MHD and HE3D ($\delta z = |\Delta^{-2}_{z_{\rm MHD}} - \Delta^{-2}_{z_{\rm HE3D}}|$; see Appendix~\ref{sec:AP_heightdiff} for an example) shows that the largest differences generally occur in the penumbra and intergranular lanes. However, the correlation between height differences and residuals of electric currents is minimal. Using the Pearson correlation coefficient as implemented in \textsc{scipy.stats.pearsonr}, we found a value of $0.06$, which shows that the errors of electric currents do not coincide with $\delta z$. 

\begin{figure}
\centering
\includegraphics[width=\linewidth]{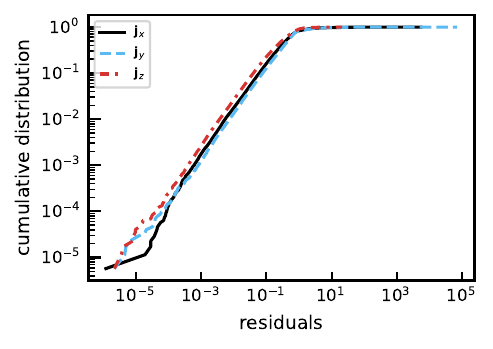}
\caption{Cumulative distribution functions for the three different components, with the relative differences between HE3D and MHD at $\mathrm{log}_{10}(\tau_{500})=-2$.}
\label{fig:main_cdf_comp}
\end{figure}

\subsection{Multiplicity of the solution}\label{subsec:shift}

\begin{figure}
    \centering
    \includegraphics[width=\linewidth]{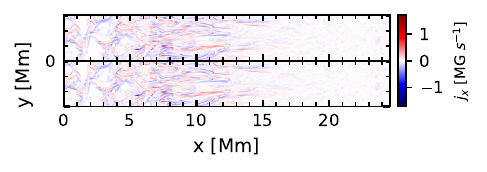}
    \caption{Map of solutions of $j_x$, where the upper map shows the solution from the solver using the unchanged input. The lower map shows the solution obtained with the shifted input.}
    \label{fig:shift_mapjx-2}
\end{figure}

Up to this point, we included a significant amount of information, which would not be given in the input to the solver in real applications. For example, the MHD input has information about a force balance that is absent when considering physical parameters from spectropolarimetric inversions. To evaluate the performance of the method on a less favorable case, we compared the $\mathrm{HE3D_{shift}}$ to the original HE3D, calculating the relative difference between the two solutions. The resulting $j_x$ for both are shown in Fig.~\ref{fig:shift_mapjx-2} at an optical depth of $\mathrm{log}_{10}\left(\tau_{500}\right)=-2$. Overall, the comparison shows a significant resemblance both in its spatial distribution and regarding their amplitudes. We found median relative differences of 0.016, 0.020, and 0.020 for $j_x$, $j_y$, and $j_z$, respectively, after excluding the masked pixels because they were below the previously mentioned noise threshold. 

\begin{figure}
    \centering
    \includegraphics[width=\linewidth]{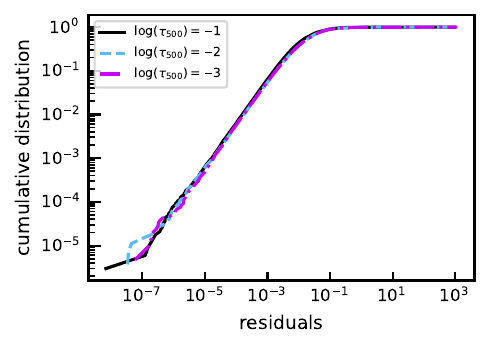}
    \caption{Cumulative distribution of relative difference between the original and the shifted solution, plotted for three different optical depths.}
    \label{fig:shift_cdf}
\end{figure}

Therefore, while the two solutions are not exactly identical, they are indeed very close. In Fig.~\ref{fig:shift_cdf}, we plot the cumulative distribution of relative differences between the two solutions for the three optical depths, where the graph shows that the residuals exhibit a scarce dependence on the optical depth. The relative difference over the full optical depth range further emphasizes this fact, where we have a median of $0.017$ for the x-component, $0.020$ for the y-component and $0.024$ for the z-component. From these numbers, we can conclude that the solution is unique, as the relative difference between the two solutions is an order of magnitude smaller than the relative difference between the solution and simulation.

\begin{figure}
    \centering
    \includegraphics[width=\linewidth]{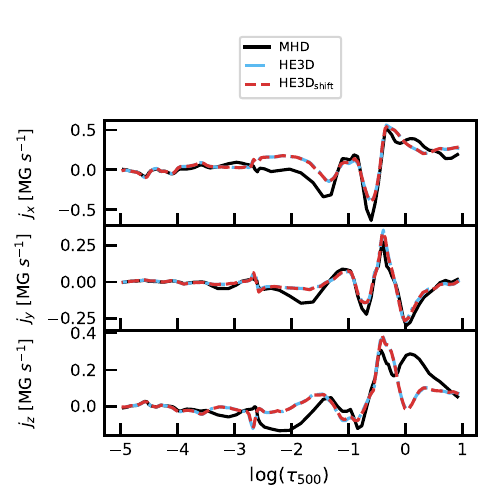}
    \caption{Electric current components for an optical depth $1>\log_{10}\left(\tau_{500}\right)>-5$ comparing both the original and shifted HE3D-run to the MHD simulation.}
    \label{fig:shift_pix_750_plsma}
\end{figure}

Figure~\ref{fig:shift_pix_750_plsma} presents the current components over the optical depth for a pixel in the penumbra (white square in Fig.~\ref{fig:map_intensity}) showing that both curves look very similar for the most part and only seem to differ significantly from each other at very few depth-points, depending on which component. Clearly, these differences are much smaller than the difference to the simulation. As this does not affect the ability of the code to reproduce the larger structure of the currents, which can be seen in the maps in Fig.~\ref{fig:shift_mapjx-2}, the solver does not appear to be affected by different starting points of the data.

\subsection{Impact of boundary condition}\label{subsec:fovsize}

\begin{figure}
    \centering
    \includegraphics[width=\linewidth]{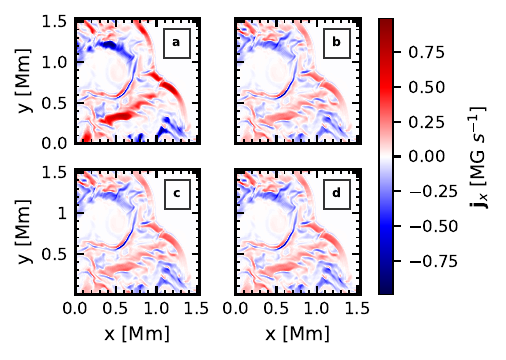}
    \caption{Maps of $j_x$ for the three different sub-fields and the simulation. The simulation is shown in a), the large FOV in b), c) represents the intermediate and d) the small FOV. All maps are at $\mathrm{log}_{10}\left(\tau_{500}\right)=-2$.}
    \label{fig:sub_map_-2}
\end{figure}

The method we developed and implemented in this study is sensitive to imposing a boundary condition on gas pressure, which we chose to be a constraint on the average pressure value for a specific iso-optical depth layer. To investigate the impact this boundary condition has on the result, we carried out two additional runs with smaller sub-fields, which are highlighted in Fig.~\ref{fig:map_intensity} by the green and cyan rectangles. The first, intermediate-sized, contains $256\times512\times256$ pixels, whereas the second, which we call the small-sized, has $128\times128\times256$ pixels. The values for the pressure reference were $9.8\cdot 10^4 \: \mathrm{dyn\: cm^{-2}} $ for the large,  $7.9\cdot 10^4 \: \mathrm{dyn\: cm^{-2}}$  for the intermediate and $7.7\cdot 10^4 \: \mathrm{dyn\: cm^{-2}}$ for the small FOV. We then restricted the analysis to the smallest field of view, the maps of $j_x$ of the three cases, and the simulation are shown in Fig.~\ref{fig:sub_map_-2} for an optical depth of $\mathrm{log}_{10}\left(\tau_{500}\right)=-2$. 

\begin{figure}
    \centering
    \includegraphics[width=\linewidth]{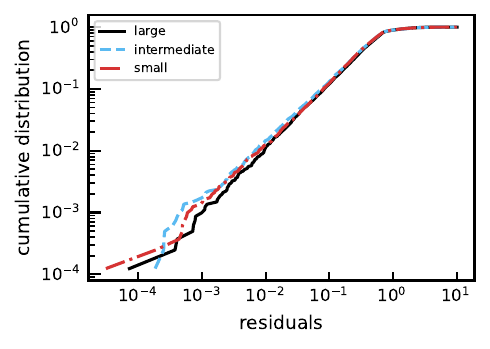}
    \caption{Cumulative distribution of relative difference of three differently sized FOVs compared to the simulation, specifically at an optical depth of $\mathrm{log}_{10}\left(\tau_{500}\right)=-2$}
    \label{fig:sub_cdf_-2}
\end{figure}

We then calculated the relative difference of these runs compared to the simulation, finding a median of $0.40$ for the large area, $0.38$ for the intermediate, and $0.38$ for the small FOV. All three cases have almost identical numbers and when looking at the maps,we can see the spatial features are also virtually indistinguishable. For the y-component, the numbers are very similar with relative differences of $0.40$ (large) and $0.38$ for both intermediate and small. In the case of ${j}_z$, we found median residuals of $0.28$ in the case of the large FOV and $0.27$ for the two smaller ones. The cumulative distribution functions of the residuals for all three areas are plotted in the graph of Fig.~\ref{fig:sub_cdf_-2}. Although they show some small differences for very small residuals, all three curves become nearly indistinguishable beyond small residuals and reach $1$ at a similar value, which means that all three chosen sub-fields perform very similarly with minimal deviations. 

Extending the analysis to the optical depth range $1>\mathrm{log}_{10}\left(\tau_{500} \right)> -5$, we found that despite the very different gas pressure and density stratifications among the three runs (not shown), which is expected due to the 3D hydrostatic equilibrium hypothesis, the retrieved electric currents have similar accuracy to the original large area that was considered in the other tests. In numbers, the smaller FOVs had a median residual of $0.25$, whereas the full area showed a median of $0.26$. The same numbers were found for the y-component. For the z-component, the median relative difference was $0.21$ for all three FOVs.

\subsection{Impact of off-centre data}\label{subsec:ocdata}

The implementation of the solver presented here assumes regular spatial sampling along the x-,y-, and z-direction (not necessarily equal in all three directions). This is suitable for the setup we consider here, but in reality, the equal sampling is done along the detector elements rather than the solar surface and along the axis chosen to invert the data (optical depth, column mass, or line of sight geometrical distance). Since the inversion of spectro-polarimetric data is usually performed on a node configuration, we could interpolate the inversion result in these nodes so that the atmosphere fulfills the regular spatial sampling required for the solver. It is also possible to reformulate the matrix, $\mathbf{A}$, so that the implementation of the derivative accounts for the actual irregular x,y,z samplings, although this approach will make the calculation of the finite difference scheme more complex, as we need to move to higher orders to achieve the same precision. Instead, here we chose to explore the consequences of ignoring the above and just assume a regular spatial sampling for an irregular one in a pessimistic real-case scenario. Rather than evaluating an individual dx and dy per pixel and using it to estimate the pixel dependent magnetic field component derivatives, we will use the two extreme values of dx and dy that we will find in a FOV size similar to that of  the CRISP-2 instrument \citep{Scharmer2026,Scharmer2008} on the Swedish Solar Telescope \citep[SST, ][]{Scharmer2003}. As this is a high resolution instrument with one of the largest FOVs, the calculation will provide one of the more extreme differences in dx and dy. We assume a viewing angle $\mu=0.6$, which is usually the maximum we should allow when applying the inversion code, as the plane-parallel atmosphere assumption will start to break when going further towards the limb. The result is that if the CRISP-instrument on the SST would have a spatial sampling of $12$ km, it will vary between $11$ and $14$ km in the extremes.

\begin{figure}
    \centering
    \includegraphics[width=\linewidth]{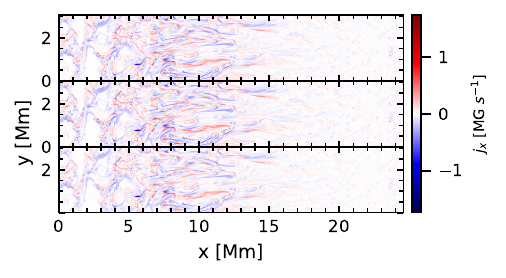}
    \caption{Map of $j_x$ for the solution with normal step size (upper), and the solutions using $11$ (middle) and $14$ (lower) km as the step size in both horizontal directions.}
    \label{fig:oc_map}
\end{figure}

Therefore, we ran the solver two times, changing the step size in both horizontal directions to $11$ and $14$ km, respectively, to account for the maximum possible difference that is to be expected. The results are then compared to the original HE3D-run, using the relative difference. The maps of the three runs with different pixel-sizes are shown in Fig.~\ref{fig:oc_map}, where the upper map shows the standard solver output, the middle shows the currents with the output using a dx/dy of $11$ km, and the lower map the same but with $14$ km. As in the previous tests, we show $j_x$ at an optical depth of $\mathrm{log}_{10} \left(\tau_{500}\right)=-2$. There, we found a median relative difference of $0.32$ for $11$ km and $0.32$ for $14$ km, compared to the $0.31$ for the HE3D-run we already showed earlier. For the y-component, the median relative difference using the $11$ km step size was $0.38$ and using $14$ km was $0.39$, compared to the $0.38$ of the original run. Finally, for the vertical component, the case of the smaller step size showed a difference of $0.30$ and  a difference of $0.34$ for the larger one; whereas the difference was $0.30$ in the original HE3D case.

\begin{figure}
    \centering
    \includegraphics[width=\linewidth]{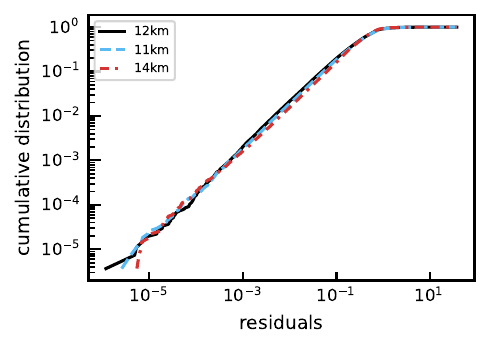}
    \caption{Cumulative distribution of relative differences between the original solver output and the output with changed step sizes.}
    \label{fig:oc_cdf}
\end{figure}

Looking at the cumulative distribution functions in Fig.~\ref{fig:oc_cdf} confirms the minimal impact of the change in step size. In the plot all three functions are almost overlapping each other, with the two off-centre cases rising slightly slower due to their higher residuals. However, these differences are small compared to the overall deviation from the simulation. 

We also looked at the median relative difference over the whole optical depth range we also used in previous cases. Starting with $j_x$, we found a median relative difference to the MHD case of $0.27$, $0.26$ for the $11$ and $14$ km cases, compared to $0.25$ for the $12$ km case. Following with $j_y$, the resulting median difference was found to be $0.27$ and $0.28$ for the two step sizes compared to $0.25$. In the case of $j_z$, the calculation resulted in median residuals of $0.28$ and $0.30$, which is almost the same as $0.26$ for the original HE3D output (see also Table~\ref{tab:summary}).

\subsection{Divergence of the magnetic field}\label{subsec:divB}

\begin{figure}
    \centering
    \includegraphics{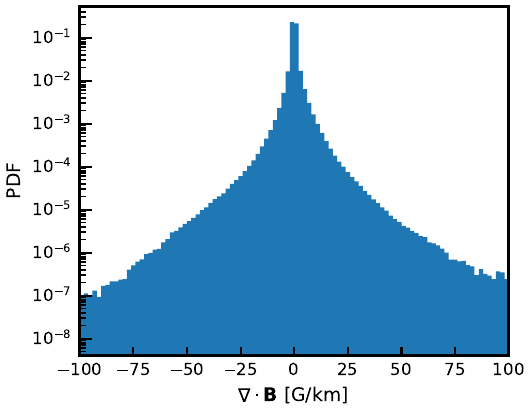}
    \caption{Probability density function of the magnetic field divergence for $\log_{10}\tau_{500}\in[1,-5]$.}
    \label{fig:divb}
\end{figure}

An additional parameter that can be estimated once we have solved the hydrostatic equilibrium in 3D is the magnetic field divergence. From a theoretical point of view the divergence of the magnetic field should vanish and it does so numerically in the simulation itself (using a fourth order centered differential scheme). However, assuming hydrostatic equilibrium in 3D to set the geometrical scale in the domain can potentially distort this property. In Fig.~\ref{fig:divb}, we present the PDF of the magnetic field divergence, which is evidently no longer 0. This is to be expected as assuming hydrostatic equilibrium in 3D involves some local stretching or compression of the physical property stratifications along the vertical direction and with that, also changes in the horizontal derivatives as the stretching or compressing will differ between pixels. Despite the long tails of the distribution, approximately 78\% (99\%) of the grid points considered here are characterised by $\nabla\cdot\boldsymbol{B}<1$ G/km ($\nabla\cdot\boldsymbol{B}<10$ G/km). As we noted previously, this encouraging result is mostly due to the fact that magnetic field divergence is a parameter with local dependency and on that scale, the assumption of hydrostatic equilibrium in 3D is not that inaccurate.

\subsection{Node-based atmosphere assumption}\label{subsec:node}

\begin{figure}[htbp]
\centering
\includegraphics[width=0.95\linewidth]{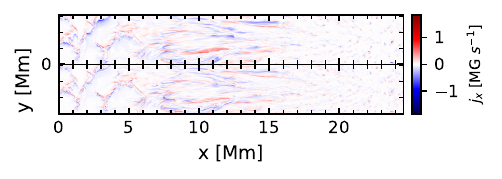}
\caption{Maps of $j_x$ for $\mathrm{MHD_{node}}$ (upper) and $\mathrm{HE3D_{node}}$ (lower). Both of the maps are at $\mathrm{log}_{10}\left(\tau_{500}\right)=-2$.}
\label{fig:main_mapjx-2_nod}
\end{figure}

While the relative difference between the MHD and HE3D cases looks promising, this is a very idealised case where we feed the solver both a temperature and magnetic field components that are very detailed in the spatial 3D domain compared to what is realistic for an inversion, where the height is represented by a few nodes. To have a better insight into this situation, we consider an ideal case with a node-sampled atmosphere, $\mathrm{MHD_{node}}$ (i.e. the difference only comes from the representation of the atmosphere with nodes). The map of ${j}_x$ for $\log_{10}\left(\tau_{500}\right)=-2$ is shown in the top graph of Fig.~\ref{fig:main_mapjx-2_nod}. In the case for the nodes, the structures seem to be smeared out, and a lot of the small-scale features are no longer present. The larger structure is still replicated though and when looking at the top maps of Fig.~\ref{fig:main_map_HE3D} and Fig.~\ref{fig:main_mapjx-2_nod}, the locations that contain the largest amplitudes are the same. However, the amplitudes themselves do not match, where the nodes representation produces much smaller peaks. This trend is visible for all three components across different optical depths, and is further reinforced by looking at the distribution of relative differences. Looking at the median relative difference at this optical depth, we found $0.74$ (x), $0.75$ (y) and $0.66$ (z) for the respective components. In general, these numbers are significantly higher than for the HE3D-run shown before. 

\begin{figure}
\centering
\includegraphics[width=0.95\linewidth]{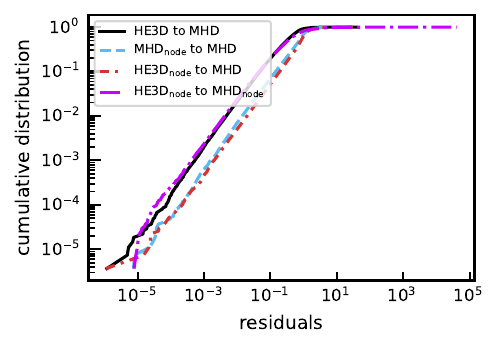}
\caption{Cumulative distribution function of relative differences between $j_x$ for different cases at $\mathrm{log}_{10}\left(\tau_{500}\right)=-2$. }
    \label{fig:cdf_jx_main}
\end{figure}

The cumulative distributions for $\log_{10}\left(\tau_{500}\right)=-2$ are plotted in Fig.~\ref{fig:cdf_jx_main}, where we also showed the difference of the previously discussed HE3D case for reference. The cumulative distributions show that the solver output is significantly closer to the simulation than the node-sampled atmosphere.

Examining the behaviour over the optical depth range shown in Fig.~\ref{fig:main_pix_1000}, it is clear that the node-sampled atmosphere can only represent a very limited amount of the structure present in the currents. Most of the strong peaks are heavily damped and a lot of the small-scale features are entirely missing, especially compared to the performance of the HE3D output. The median relative difference over the range of $1>\log_{10}\left(\tau_{500}\right)>-5$ was found to be $0.60$, $0.68$, and $0.63$ for $j_x$, $j_y$, and $j_z$ respectively, which are also consistently higher than for HE3D. 

Although the number of nodes employed will have a clear impact, it is evident from these results that the currents obtained after assuming a node-based model representation are different. That is the reason why we 
investigate the impact of the method ($\mathrm{HE3D_{node}}$) implemented in this work, on these interpolated parameters as input for the solver. The map of $j_x$ is shown on the lower panel of Fig.~\ref{fig:main_mapjx-2_nod} for the same optical depth of $\mathrm{log}_{10}\left(\tau_{500} \right)=-2$. In Fig.~\ref{fig:cdf_jx_main}, the cumulative distribution of relative errors compared to the simulation, but also the difference between the node sampled atmosphere and $\mathrm{HE3D_{node}}$ (i.e. output versus input) is shown. 

We can see that the output using the interpolated input reproduces the structure quite well, although it naturally shows the same smearing as well as the significantly lower amplitudes as the input map above. A look at the cumulative distribution confirms this, as $\mathrm{HE3D_{node}}$ performs only slightly worse than $\mathrm{MHD_{node}}$ when comparing to the simulation. Additionally, when calculating the relative difference between the two, marked in the plot by the violet curve, we can see that it is very similar to the original solver output HE3D to MHD comparison (black). The latter is shown by looking at the median value for the different cases: for the comparison to MHD, we find values of $0.84$, $0.85$ and $0.76$ for $j_x$, $j_y$ and $j_z$. If we compare them to the node-sampled input, the relative differences are $0.34$ (x), $0.41$ (y), and $0.36$ (z).
Following up by looking at the behaviour in Fig.~\ref{fig:main_pix_1000}, we see that $\mathrm{HE3D_{node}}$ is close to the node-sampled MHD-atmosphere, but suffers the same shortcomings. Compared to the MHD-case, the small-scale features are now missing, but the large-scale structure is successfully reproduced. Comparing to $\mathrm{MHD_{node}}$, the difference is reasonably small, and the areas with the largest current amplitudes are the same. 
We also looked at the median relative difference of the data cube within the optical depth range, where we found a difference of $0.69$, $0.76$, and $0.70$ for x-,y-, and z-component. In doing the comparison with the node-based atmosphere input, we find numbers of $0.25$ for $j_x$, $0.27$ for $j_y$, and $0.28$ for $j_z$. From these results, we conclude that the method performs similarly when provided with either a node-based atmospheric representation or a much more spatially detailed input.

\section{Discussion}\label{sec:4}

Table~\ref{tab:summary} shows a summary of the median relative error values from the various runs presented above. The results show promise, as the difference is low enough to allow us to obtain good constraints on the distribution of heating terms based on $\boldsymbol{j}$. Although the amplitudes of the peaks are off compared to the simulation, the regions that contain the strongest currents are the same, which allows us to determine the areas where heating takes place and provide lower limit constraints on dissipative heating rates.
In possible future applications of the presented method to observational data, the ground truth will be unknown. Therefore, the mask to exclude noise can only be applied to the end result of the solver. To test the impact of this, we also calculated the error with the mask determined by the HE3D case numbers, not MHD. The median values at $\mathrm{log}_{10}(\tau_{500})=-2$ were $0.35$, $0.40$ and $0.28$. Over the optical depth range of $\log\tau_{500} [1,-5]$, the median residuals were $0.27$, $0.32$ and $0.26$, which are very similar to the ones in Table~\ref{tab:summary} (i.e. the statistical performance is still similar).

We also saw that the method performed similarly well for configurations aiming to represent realistic cases more closely, such as the lower depth resolution used in inversion codes. The main factor that sets the accuracy of the reconstruction of the electric current density vector is the mapping between optical depth to geometrical distance, which does actually come from the assumption of hydrostatic equilibrium in 3D. This relation appears due to the fact that the spectra are sensitive to the various physical parameters in optical depth and its transformation to geometrical distance comes from absorption occurring along the path traveled by radiation. 

\begin{table}[]
    \centering
    \caption{All median relative differences both at an optical depth of $\mathrm{log}_{10}\left(\tau_{500}\right)=-2$ and for the entire range of $\log_{10}(\tau_{500})\in [1,-5]$ in brackets.}
    \scriptsize
    \begin{tabular}{|c|c|c|c|c|}
    \hline
        case & reference & ${j}_x$ & ${j}_y$ & ${j}_z$\\
        \hhline{|=|=|=|=|=|}
        HE3D & MHD & $0.31$ ($0.25$) & $0.38$ ($0.25$)& $0.30$ ($0.26$)\\
        $\mathrm{MHD_{nodes}}$ & MHD & $0.74$ ($0.60$) & $0.75$ ($0.68$)& $0.66$ ($0.63$)\\
        $\mathrm{HE3D_{nodes}}$ & MHD & $0.84$ ($0.69$) & $0.85$ ($0.76$)& $0.76$ ($0.70$)\\
        $\mathrm{HE3D_{nodes}}$ & $\mathrm{MHD_{nodes}}$ & $0.34$ ($0.25$) & $0.41$ ($0.27$)& $0.36$ ($0.28$)\\
        \hline
        $\mathrm{HE3D_{shift}}$ & HE3D & $0.016$ ($0.017$) & $0.020$ ($0.020$)& $0.020$ ($0.024$)\\
        \hline
        $\mathrm{HE3D_{large}}$ & MHD & $0.40$ ($0.26$) & $0.40$ ($0.26$)& $0.28$ ($0.21$)\\
        $\mathrm{HE3D_{intermediate}}$ & MHD & $0.38$ ($0.25$) & $0.38$ ($0.25$)& $0.27$ ($0.21$)\\
        $\mathrm{HE3D_{small}}$ & MHD & $0.38$ ($0.25$) & $0.38$ ($0.25$)& $0.27$ ($0.21$)\\
        \hline
        $\mathrm{HE3D_{11 km}}$ & MHD & $0.32$ ($0.27$) & $0.38$ ($0.27$)& $0.30$ ($0.28$)\\
        $\mathrm{HE3D_{14 km}}$ & MHD & $0.32$ ($0.26$) & $0.39$ ($0.28$)& $0.34$ ($0.30$)\\
    \hline
    \end{tabular}
    \label{tab:summary}
\end{table}

Although we have not considered radiation explicitly in this work, it has entered into the method as we impose that temperature has to be kept the same at each optical depth point. In doing so, we enforce that the spectra synthesized from any of the runs is the same. This is the main reason behind the differences we determined in Sect.~\ref{sect:he3d_ideal}. In that case, the only reason for the difference in the electric current density estimation comes from the local expansion and compression of the physical parameters stratification associated with the change in how the opacity occurs. The fact that this expansion and contraction (at least in the optical depth range between $\log\tau_{500} [1,-5]$) is not extreme is the reason why this method achieves such acceptable electric current density estimation. This surprising result is due to the fact that the electric current density calculation depends on physical parameters only locally, and that is why this method is reasonably good at estimating electric current densities. From Fig.~\ref{fig:main_hist_z}, we can also conclude that the current densities are generally placed at the correct relative heights, with only small differences. Similarly, we have seen that the assumption of hydrostatic equilibrium in 3D induces a non-zero magnetic field divergence (as compared to the simulation), but the deviations are small with more than 70\% of the grid points with $\nabla\cdot\mathrm{B}<1$ G/km. As for the electric current density estimation, this is possible due to the local dependency of this parameter on the magnetic field vector. However, if we consider other parameters that are global in nature (such as the Wilson depression; see e.g. in Fig.~\ref{fig:AP_main_wd0}), then this method cannot retrieve any meaningfully acceptable result, as it strongly modifies the pressure balance over the entire volume. 

One of the major advantages of this method is that it suffers little impact from external constraints. It does not rely on disambiguating the magnetic field vector, which is still challenging to perform for the chromosphere and it does not require having the magnetic field inferred for large FOV. Moreover, it relies on a single boundary condition that has little impact on the result, regardless of the size of the FOV or its boundaries. Finally, the method converges to equivalent solutions independently of the initial state.

We validated these results over the range of $\log_{10}\tau_{500} [1,-5]$, meaning that this method potentially enables the inference of electric current densities from the photosphere to the chromosphere. We have seen that the method recovers a magnetic-field divergence close to zero without using the magnetic field itself to determine the geometry. This property could be used to further constrain the magnetic field 180$^{\circ}$ ambiguity \citep[][and references therein]{metcalf2006}. More generally, the ability to generally extract the distances between optical depth layers with reasonable accuracy could also prove useful in other applications for determining the geometrical locations of other physical phenomena such as enhanced radiative losses events. 

With the application of the method to observational data in mind, we evaluated the effect of non-uniform step size introduced when analyzing data taken away from the disc centre and found that it has less of an impact than other sources of uncertainty or limitations tested here. It is important to emphasize that even though we have evaluated the performance of the method in some set-ups that aimed at mimicking realistic scenarios, its future application to observational data ought to consider additional instrument-specific effects. The accuracy of the reconstructed atmospheric parameters will depend on the instrument and observational setup, including the selected spectral lines, integration time, spectral and spatial resolution, and instrumental systematics. Data-reduction procedures and inversion choices will introduce further uncertainties in the atmospheric parameters used as input to the method. It is therefore unrealistic to expect a similar accuracy to what is shown in this work with the input from the simulation. As a final remark, while the method does need significant computational expense on its own, its computational cost is several orders of magnitude lower than what is needed by the inversion code. 

\section{Conclusions and outlook} \label{sec:5}

In this paper, we present a method to solve the hydrostatic equilibrium equations in 3D in a Cartesian grid. We developed this solver to be able to estimate electric currents, which require the calculation of spatial derivatives of the magnetic field. Specifically, we designed it to be applied on the inferred physical parameters of depth-stratified spectropolarimetric inversion codes as a final post-processing step. We used a snapshot from an r-MHD simulation to test the validity of our approach. 

Starting with the general performance of the method, we found that the method is able to retrieve around 50\% of the grid points with less than 30\% of relative difference. Furthermore, it performs similarly in cases where the amount of information in the vertical stratification of the physical properties is reduced. This case is more similar to what common depth-stratified inversion codes infer. We also ascertained that the solution is unique (within numerical precision), regardless of the amount of spatial information contained in the input (i.e. if the input has horizontal force balance or not), as well as the impact of using different boundary conditions for the gas pressure. Furthermore, we have tested real-world scenarios, both addressing the effect of having observations out of the disc centre and the usage of nodes to represent the state of the atmosphere (most common approach in spectropolarimetric observations). 
 
Overall, these more realistic applications maintain a similar performance of the implemented method, thereby promising the determination of electric current densities (lower boundaries) and, by extension, estimating lower limits of dissipative heating terms. The code will now be implemented into the STockholm inversion Code STiC \citep{delaCruzRodriguez2019b,delaCruzRodriguez2019a} as an optional module, where we plan to apply it to spectropolarimetric observations of the chromosphere and use the final output to estimate the ohmic heating and ambipolar diffusion terms. While we successfully tested the code assuming hydrostatic equilibrium in three dimensions, in a future paper, we will consider implementing other approximations that would allow us to infer additional information.

\begin{acknowledgements}
The Institute for Solar Physics is supported by a grant for research infrastructures of national importance from the Swedish Research Council (registration number 2021-00169). 
APY acknowledges support from the Swedish Research Council (grant 2023-03313). This project has been funded by the European Union through the European Research Council (ERC) under the Horizon Europe program (MAGHEAT, grant agreement 101088184). 
We also thank Mathias Rempel for making his simulation publicly available and the volunteers that keep developing the Eigen library.
We also wish to thank the referee for the comments and suggestions to improve this work.
\end{acknowledgements}

\bibliographystyle{aa}
\bibliography{bibli}

@ARTICLE{PastorYabar2019,
       author = {{Pastor Yabar}, A. and {Borrero}, J.~M. and {Ruiz Cobo}, B.},
        title = "{FIRTEZ-dz. A forward and inverse solver of the polarized radiative transfer equation under Zeeman regime in geometrical scale}",
      journal = {\aap},
         year = 2019,
        month = sep,
       volume = {629},
          eid = {A24},
        pages = {A24},
          doi = {10.1051/0004-6361/201935692},
archivePrefix = {arXiv},
       eprint = {1908.08075},
 primaryClass = {astro-ph.SR},
       adsurl = {https://ui.adsabs.harvard.edu/abs/2019A&A...629A..24P}
}

@ARTICLE{delaCruzRodriguez2019a,
       author = {{de la Cruz Rodr{\'\i}guez}, J.},
        title = "{A method for global inversion of multi-resolution solar data}",
      journal = {\aap},
         year = 2019,
        month = nov,
       volume = {631},
          eid = {A153},
        pages = {A153},
          doi = {10.1051/0004-6361/201936635},
archivePrefix = {arXiv},
       eprint = {1909.02604},
 primaryClass = {astro-ph.SR},
       adsurl = {https://ui.adsabs.harvard.edu/abs/2019A&A...631A.153D}
}

@ARTICLE{delaCruzRodriguez2019b,
       author = {{de la Cruz Rodr{\'\i}guez}, J. and {Leenaarts}, J. and {Danilovic}, S. and {Uitenbroek}, H.},
        title = "{STiC: A multiatom non-LTE PRD inversion code for full-Stokes solar observations}",
      journal = {\aap},
         year = 2019,
        month = mar,
       volume = {623},
          eid = {A74},
        pages = {A74},
          doi = {10.1051/0004-6361/201834464},
archivePrefix = {arXiv},
       eprint = {1810.08441},
 primaryClass = {astro-ph.SR},
       adsurl = {https://ui.adsabs.harvard.edu/abs/2019A&A...623A..74D}
}

@ARTICLE{delToroIniesta2016,
       author = {{del Toro Iniesta}, Jose Carlos and {Ruiz Cobo}, Basilio},
        title = "{Inversion of the radiative transfer equation for polarized light}",
      journal = {Living Reviews in Solar Physics},
         year = 2016,
        month = dec,
       volume = {13},
       number = {1},
          eid = {4},
        pages = {4},
          doi = {10.1007/s41116-016-0005-2},
archivePrefix = {arXiv},
       eprint = {1610.10039},
 primaryClass = {astro-ph.SR},
       adsurl = {https://ui.adsabs.harvard.edu/abs/2016LRSP...13....4D}
}

@article{Nobregar_Siverio_et_al2020,
	author = {{Nóbrega-Siverio, D.} and {Moreno-Insertis, F.} and {Martínez-Sykora, J.} and {Carlsson, M.} and {Szydlarski, M.}},
	title = {Nonequilibrium ionization and ambipolar diffusion in solar magnetic flux emergence processes},
	DOI= "10.1051/0004-6361/201936944",
	url= "https://doi.org/10.1051/0004-6361/201936944",
	journal = {\aap},
	year = 2020,
	volume = 633,
	pages = "A66",
}

@ARTICLE{Milic_et_al2018,
       author = {{Mili{\'c}}, I. and {van Noort}, M.},
        title = "{Spectropolarimetric NLTE inversion code SNAPI}",
      journal = {\aap},
         year = 2018,
        month = sep,
       volume = {617},
          eid = {A24},
        pages = {A24},
          doi = {10.1051/0004-6361/201833382},
archivePrefix = {arXiv},
       eprint = {1806.08134},
 primaryClass = {astro-ph.SR},
       adsurl = {https://ui.adsabs.harvard.edu/abs/2018A&A...617A..24M}
}

@article{Rempel_2012,
   title={NUMERICAL SUNSPOT MODELS: ROBUSTNESS OF PHOTOSPHERIC VELOCITY AND MAGNETIC FIELD STRUCTURE},
   volume={750},
   ISSN={1538-4357},
   url={http://dx.doi.org/10.1088/0004-637X/750/1/62},
   DOI={10.1088/0004-637x/750/1/62},
   number={1},
   journal={The Astrophysical Journal},
   publisher={American Astronomical Society},
   author={Rempel, M.},
   year={2012},
   month=apr, pages={62} }

@article{vanderHorst_1992,
author = {van der Vorst, H. A.},
title = {Bi-CGSTAB: A Fast and Smoothly Converging Variant of Bi-CG for the Solution of Nonsymmetric Linear Systems},
journal = {SIAM Journal on Scientific and Statistical Computing},
volume = {13},
number = {2},
pages = {631-644},
year = {1992},
doi = {10.1137/0913035},
URL = {https://doi.org/10.1137/0913035},
eprint = {https://doi.org/10.1137/0913035}
}

@ARTICLE{Voegler2005,
       author = {{V{\"o}gler}, A. and {Shelyag}, S. and {Sch{\"u}ssler}, M. and {Cattaneo}, F. and {Emonet}, T. and {Linde}, T.},
        title = "{Simulations of magneto-convection in the solar photosphere.  Equations, methods, and results of the MURaM code}",
      journal = {\aap},
         year = 2005,
        month = jan,
       volume = {429},
        pages = {335-351},
          doi = {10.1051/0004-6361:20041507},
       adsurl = {https://ui.adsabs.harvard.edu/abs/2005A&A...429..335V}
}

@ARTICLE{Louis2021,
       author = {{Louis}, Rohan E. and {Prasad}, Avijeet and {Beck}, Christian and {Choudhary}, Debi P. and {Yalim}, Mehmet S.},
        title = "{Heating of the solar chromosphere in a sunspot light bridge by electric currents}",
      journal = {\aap},
         year = 2021,
        month = aug,
       volume = {652},
          eid = {L4},
        pages = {L4},
          doi = {10.1051/0004-6361/202141456},
archivePrefix = {arXiv},
       eprint = {2107.12066},
 primaryClass = {astro-ph.SR},
       adsurl = {https://ui.adsabs.harvard.edu/abs/2021A&A...652L...4L}
}

@ARTICLE{Khomenko2012,
       author = {{Khomenko}, E. and {Collados}, M.},
        title = "{Heating of the Magnetized Solar Chromosphere by Partial Ionization Effects}",
      journal = {\apj},
         year = 2012,
        month = mar,
       volume = {747},
       number = {2},
          eid = {87},
        pages = {87},
          doi = {10.1088/0004-637X/747/2/87},
archivePrefix = {arXiv},
       eprint = {1112.3374},
 primaryClass = {astro-ph.SR},
       adsurl = {https://ui.adsabs.harvard.edu/abs/2012ApJ...747...87K}
}

@ARTICLE{Shelyag2016,
       author = {{Shelyag}, S. and {Khomenko}, E. and {de Vicente}, A. and {Przybylski}, D.},
        title = "{Heating of the Partially Ionized Solar Chromosphere by Waves in Magnetic Structures}",
      journal = {\apjl},
         year = 2016,
        month = mar,
       volume = {819},
       number = {1},
          eid = {L11},
        pages = {L11},
          doi = {10.3847/2041-8205/819/1/L11},
archivePrefix = {arXiv},
       eprint = {1602.03373},
 primaryClass = {astro-ph.SR},
       adsurl = {https://ui.adsabs.harvard.edu/abs/2016ApJ...819L..11S}
}

@INPROCEEDINGS{Scharmer2003,
       author = {{Scharmer}, Goran B. and {Bjelksjo}, Klas and {Korhonen}, Tapio K. and {Lindberg}, Bo and {Petterson}, Bertil},
        title = "{The 1-meter Swedish solar telescope}",
    booktitle = {Innovative Telescopes and Instrumentation for Solar Astrophysics},
         year = 2003,
       editor = {{Keil}, Stephen L. and {Avakyan}, Sergey V.},
       series = {Society of Photo-Optical Instrumentation Engineers (SPIE) Conference Series},
       volume = {4853},
        month = feb,
        pages = {341-350},
          doi = {10.1117/12.460377},
       adsurl = {https://ui.adsabs.harvard.edu/abs/2003SPIE.4853..341S}
}

@ARTICLE{Scharmer2008,
       author = {{Scharmer}, G.~B. and {Narayan}, G. and {Hillberg}, T. and {de la Cruz Rodriguez}, J. and {L{\"o}fdahl}, M.~G. and {Kiselman}, D. and {S{\"u}tterlin}, P. and {van Noort}, M. and {Lagg}, A.},
        title = "{CRISP Spectropolarimetric Imaging of Penumbral Fine Structure}",
      journal = {\apjl},
         year = 2008,
        month = dec,
       volume = {689},
       number = {1},
        pages = {L69},
          doi = {10.1086/595744},
archivePrefix = {arXiv},
       eprint = {0806.1638},
 primaryClass = {astro-ph},
       adsurl = {https://ui.adsabs.harvard.edu/abs/2008ApJ...689L..69S}
}

@BOOK{Athay1976,
       author = {{Athay}, R.~G.},
        title = "{The solar chromosphere and corona: Quiet sun}",
         year = 1976,
       volume = {53},
          doi = {10.1007/978-94-010-1715-2},
       adsurl = {https://ui.adsabs.harvard.edu/abs/1976ASSL...53.....A}
}

@ARTICLE{Withbroe1977,
       author = {{Withbroe}, G.~L. and {Noyes}, R.~W.},
        title = "{Mass and energy flow in the solar chromosphere and corona.}",
      journal = {\araa},
         year = 1977,
        month = jan,
       volume = {15},
        pages = {363-387},
          doi = {10.1146/annurev.aa.15.090177.002051},
       adsurl = {https://ui.adsabs.harvard.edu/abs/1977ARA&A..15..363W}
}

@ARTICLE{DiazBaso2021,
       author = {{D{\'\i}az Baso}, C.~J. and {de la Cruz Rodr{\'\i}guez}, J. and {Leenaarts}, J.},
        title = "{An observationally constrained model of strong magnetic reconnection in the solar chromosphere. Atmospheric stratification and estimates of heating rates}",
      journal = {\aap},
         year = 2021,
        month = mar,
       volume = {647},
          eid = {A188},
        pages = {A188},
          doi = {10.1051/0004-6361/202040111},
archivePrefix = {arXiv},
       eprint = {2012.06229},
 primaryClass = {astro-ph.SR},
       adsurl = {https://ui.adsabs.harvard.edu/abs/2021A&A...647A.188D}
}

@ARTICLE{Srivastava2021,
       author = {{Srivastava}, A.~K. and {Ballester}, J.~L. and {Cally}, P.~S. and {Carlsson}, M. and {Goossens}, M. and {Jess}, D.~B. and {Khomenko}, E. and {Mathioudakis}, M. and {Murawski}, K. and {Zaqarashvili}, T.~V.},
        title = "{Chromospheric Heating by Magnetohydrodynamic Waves and Instabilities}",
      journal = {Journal of Geophysical Research (Space Physics)},
         year = 2021,
        month = jun,
       volume = {126},
       number = {6},
          eid = {e029097},
        pages = {e029097},
          doi = {10.1029/2020JA029097},
archivePrefix = {arXiv},
       eprint = {2104.02010},
 primaryClass = {astro-ph.SR},
       adsurl = {https://ui.adsabs.harvard.edu/abs/2021JGRA..12629097S}
}

@ARTICLE{Abbasvand2020,
       author = {{Abbasvand}, V. and {Sobotka}, M. and {{\v{S}}vanda}, M. and {Heinzel}, P. and {Garc{\'\i}a-Rivas}, M. and {Denker}, C. and {Balthasar}, H. and {Verma}, M. and {Kontogiannis}, I. and {Koza}, J. and {Korda}, D. and {Kuckein}, C.},
        title = "{Observational study of chromospheric heating by acoustic waves}",
      journal = {\aap},
         year = 2020,
        month = oct,
       volume = {642},
          eid = {A52},
        pages = {A52},
          doi = {10.1051/0004-6361/202038559},
archivePrefix = {arXiv},
       eprint = {2008.02688},
 primaryClass = {astro-ph.SR},
       adsurl = {https://ui.adsabs.harvard.edu/abs/2020A&A...642A..52A}
}

@ARTICLE{SocasNavarro2005,
       author = {{Socas-Navarro}, H.},
        title = "{Are Electric Currents Heating the Magnetic Chromosphere?}",
      journal = {\apjl},
         year = 2005,
        month = nov,
       volume = {633},
       number = {1},
        pages = {L57-L60},
          doi = {10.1086/498145},
archivePrefix = {arXiv},
       eprint = {astro-ph/0508689},
 primaryClass = {astro-ph},
       adsurl = {https://ui.adsabs.harvard.edu/abs/2005ApJ...633L..57S}
}

@ARTICLE{Tritschler2008,
       author = {{Tritschler}, A. and {Uitenbroek}, H. and {Reardon}, K.},
        title = "{Evidence for a Current Sheet above a Sunspot Umbra}",
      journal = {\apjl},
         year = 2008,
        month = oct,
       volume = {686},
       number = {1},
        pages = {L45},
          doi = {10.1086/592817},
       adsurl = {https://ui.adsabs.harvard.edu/abs/2008ApJ...686L..45T}
}

@ARTICLE{SocasNavarro2015,
       author = {{Socas-Navarro}, H. and {de la Cruz Rodr{\'\i}guez}, J. and {Asensio Ramos}, A. and {Trujillo Bueno}, J. and {Ruiz Cobo}, B.},
        title = "{An open-source, massively parallel code for non-LTE synthesis and inversion of spectral lines and Zeeman-induced Stokes profiles}",
      journal = {\aap},
         year = 2015,
        month = may,
       volume = {577},
          eid = {A7},
        pages = {A7},
          doi = {10.1051/0004-6361/201424860},
archivePrefix = {arXiv},
       eprint = {1408.6101},
 primaryClass = {astro-ph.SR},
       adsurl = {https://ui.adsabs.harvard.edu/abs/2015A&A...577A...7S}
}

@ARTICLE{RuizCobo2022,
       author = {{Ruiz Cobo}, B. and {Quintero Noda}, C. and {Gafeira}, R. and {Uitenbroek}, H. and {Orozco Su{\'a}rez}, D. and {P{\'a}ez Ma{\~n}{\'a}}, E.},
        title = "{DeSIRe: Departure coefficient aided Stokes Inversion based on Response functions}",
      journal = {\aap},
         year = 2022,
        month = apr,
       volume = {660},
          eid = {A37},
        pages = {A37},
          doi = {10.1051/0004-6361/202140877},
archivePrefix = {arXiv},
       eprint = {2202.02226},
 primaryClass = {astro-ph.SR},
       adsurl = {https://ui.adsabs.harvard.edu/abs/2022A&A...660A..37R}
}

@ARTICLE{borrero2019,
       author = {{Borrero}, J.~M. and {Pastor Yabar}, A. and {Rempel}, M. and {Ruiz Cobo}, B.},
        title = "{Combining magnetohydrostatic constraints with Stokes profiles inversions. I. Role of boundary conditions}",
      journal = {\aap},
         year = 2019,
        month = dec,
       volume = {632},
          eid = {A111},
        pages = {A111},
          doi = {10.1051/0004-6361/201936367},
archivePrefix = {arXiv},
       eprint = {1910.14131},
 primaryClass = {astro-ph.SR},
       adsurl = {https://ui.adsabs.harvard.edu/abs/2019A&A...632A.111B}
}

@ARTICLE{Leenaarts2020,
       author = {{Leenaarts}, Jorrit},
        title = "{Radiation hydrodynamics in simulations of the solar atmosphere}",
      journal = {Living Reviews in Solar Physics},
         year = 2020,
        month = dec,
       volume = {17},
       number = {1},
          eid = {3},
        pages = {3},
          doi = {10.1007/s41116-020-0024-x},
archivePrefix = {arXiv},
       eprint = {2002.03623},
 primaryClass = {astro-ph.SR},
       adsurl = {https://ui.adsabs.harvard.edu/abs/2020LRSP...17....3L}
}

@ARTICLE{Carlsson2019,
       author = {{Carlsson}, Mats and {De Pontieu}, Bart and {Hansteen}, Viggo H.},
        title = "{New View of the Solar Chromosphere}",
      journal = {\araa},
         year = 2019,
        month = aug,
       volume = {57},
        pages = {189-226},
          doi = {10.1146/annurev-astro-081817-052044},
       adsurl = {https://ui.adsabs.harvard.edu/abs/2019ARA&A..57..189C}
}

@ARTICLE{Lagg2017,
       author = {{Lagg}, Andreas and {Lites}, Bruce and {Harvey}, Jack and {Gosain}, Sanjay and {Centeno}, Rebecca},
        title = "{Measurements of Photospheric and Chromospheric Magnetic Fields}",
      journal = {\ssr},
         year = 2017,
        month = sep,
       volume = {210},
       number = {1-4},
        pages = {37-76},
          doi = {10.1007/s11214-015-0219-y},
archivePrefix = {arXiv},
       eprint = {1510.06865},
 primaryClass = {astro-ph.SR},
       adsurl = {https://ui.adsabs.harvard.edu/abs/2017SSRv..210...37L}
}

@ARTICLE{Pontin2022,
       author = {{Pontin}, David I. and {Priest}, Eric R.},
        title = "{Magnetic reconnection: MHD theory and modelling}",
      journal = {Living Reviews in Solar Physics},
         year = 2022,
        month = dec,
       volume = {19},
       number = {1},
          eid = {1},
        pages = {1},
          doi = {10.1007/s41116-022-00032-9},
       adsurl = {https://ui.adsabs.harvard.edu/abs/2022LRSP...19....1P}
}

@ARTICLE{Morton2012,
       author = {{Morton}, Richard J. and {Verth}, Gary and {Jess}, David B. and {Kuridze}, David and {Ruderman}, Michael S. and {Mathioudakis}, Mihalis and {Erd{\'e}lyi}, Robertus},
        title = "{Observations of ubiquitous compressive waves in the Sun's chromosphere}",
      journal = {Nature Communications},
         year = 2012,
        month = dec,
       volume = {3},
          eid = {1315},
        pages = {1315},
          doi = {10.1038/ncomms2324},
archivePrefix = {arXiv},
       eprint = {1306.4124},
 primaryClass = {astro-ph.SR},
       adsurl = {https://ui.adsabs.harvard.edu/abs/2012NatCo...3.1315M}
}

@ARTICLE{delaCruzRodriguez2013,
       author = {{de la Cruz Rodr{\'\i}guez}, J. and {De Pontieu}, B. and {Carlsson}, M. and {Rouppe van der Voort}, L.~H.~M.},
        title = "{Heating of the Magnetic Chromosphere: Observational Constraints from Ca II {\ensuremath{\lambda}}8542 Spectra}",
      journal = {\apjl},
         year = 2013,
        month = feb,
       volume = {764},
       number = {1},
          eid = {L11},
        pages = {L11},
          doi = {10.1088/2041-8205/764/1/L11},
archivePrefix = {arXiv},
       eprint = {1301.3141},
 primaryClass = {astro-ph.SR},
       adsurl = {https://ui.adsabs.harvard.edu/abs/2013ApJ...764L..11D}
}

@ARTICLE{Brady2016,
       author = {{Brady}, C.~S. and {Arber}, T.~D.},
        title = "{Simulations of Alfv{\'e}n and Kink Wave Driving of the Solar Chromosphere: Efficient Heating and Spicule Launching}",
      journal = {\apj},
         year = 2016,
        month = oct,
       volume = {829},
       number = {2},
          eid = {80},
        pages = {80},
          doi = {10.3847/0004-637X/829/2/80},
archivePrefix = {arXiv},
       eprint = {1601.07835},
 primaryClass = {astro-ph.SR},
       adsurl = {https://ui.adsabs.harvard.edu/abs/2016ApJ...829...80B}
}

@ARTICLE{Vernazza1981,
       author = {{Vernazza}, J.~E. and {Avrett}, E.~H. and {Loeser}, R.},
        title = "{Structure of the solar chromosphere. III. Models of the EUV brightness components of the quiet sun.}",
      journal = {\apjs},
         year = 1981,
        month = apr,
       volume = {45},
        pages = {635-725},
          doi = {10.1086/190731},
       adsurl = {https://ui.adsabs.harvard.edu/abs/1981ApJS...45..635V}
}

@MISC{eigenweb,
  author = {Ga\"{e}l Guennebaud and Beno\^{i}t Jacob and others},
  title = {Eigen},
  howpublished = {https://libeigen.gitlab.io},
  year = {2010}
 }

@ARTICLE{delaCruz2017,
       author = {{de la Cruz Rodr{\'\i}guez}, J. and {van Noort}, M.},
        title = "{Radiative Diagnostics in the Solar Photosphere and Chromosphere}",
      journal = {\ssr},
         year = 2017,
        month = sep,
       volume = {210},
       number = {1-4},
        pages = {109-143},
          doi = {10.1007/s11214-016-0294-8},
archivePrefix = {arXiv},
       eprint = {1609.08324},
 primaryClass = {astro-ph.SR},
       adsurl = {https://ui.adsabs.harvard.edu/abs/2017SSRv..210..109D}
}

@ARTICLE{Li2022,
       author = {{Li}, H. and {del Pino Alem{\'a}n}, T. and {Trujillo Bueno}, J. and {Casini}, R.},
        title = "{TIC: A Stokes Inversion Code for Scattering Polarization with Partial Frequency Redistribution and Arbitrary Magnetic Fields}",
      journal = {\apj},
         year = 2022,
        month = jul,
       volume = {933},
       number = {2},
          eid = {145},
        pages = {145},
          doi = {10.3847/1538-4357/ac745c},
archivePrefix = {arXiv},
       eprint = {2205.15666},
 primaryClass = {astro-ph.SR},
       adsurl = {https://ui.adsabs.harvard.edu/abs/2022ApJ...933..145L}
}

@ARTICLE{2011ApJ...736....3V,
       author = {{van Ballegooijen}, A.~A. and {Asgari-Targhi}, M. and {Cranmer}, S.~R. and {DeLuca}, E.~E.},
        title = "{Heating of the Solar Chromosphere and Corona by Alfv{\'e}n Wave Turbulence}",
      journal = {\apj},
         year = 2011,
        month = jul,
       volume = {736},
       number = {1},
          eid = {3},
        pages = {3},
          doi = {10.1088/0004-637X/736/1/3},
archivePrefix = {arXiv},
       eprint = {1105.0402},
 primaryClass = {astro-ph.SR},
       adsurl = {https://ui.adsabs.harvard.edu/abs/2011ApJ...736....3V}
}

@ARTICLE{2010ApJ...720.1417P,
       author = {{Puschmann}, K.~G. and {Ruiz Cobo}, B. and {Mart{\'\i}nez Pillet}, V.},
        title = "{A Geometrical Height Scale for Sunspot Penumbrae}",
      journal = {\apj},
         year = 2010,
        month = sep,
       volume = {720},
       number = {2},
        pages = {1417-1431},
          doi = {10.1088/0004-637X/720/2/1417},
archivePrefix = {arXiv},
       eprint = {1007.2779},
 primaryClass = {astro-ph.SR},
       adsurl = {https://ui.adsabs.harvard.edu/abs/2010ApJ...720.1417P}
}

@ARTICLE{Metcalf2006,
       author = {{Metcalf}, Thomas R. and {Leka}, K.~D. and {Barnes}, Graham and {Lites}, Bruce W. and {Georgoulis}, Manolis K. and {Pevtsov}, A.~A. and {Balasubramaniam}, K.~S. and {Gary}, G. Allen and {Jing}, Ju and {Li}, Jing and {Liu}, Y. and {Wang}, H.~N. and {Abramenko}, Valentyna and {Yurchyshyn}, Vasyl and {Moon}, Y.-J.},
        title = "{An Overview of Existing Algorithms for Resolving the 180$^{{\textdegree}}$ Ambiguity in Vector Magnetic Fields: Quantitative Tests with Synthetic Data}",
      journal = {\solphys},
         year = 2006,
        month = sep,
       volume = {237},
       number = {2},
        pages = {267-296},
          doi = {10.1007/s11207-006-0170-x},
       adsurl = {https://ui.adsabs.harvard.edu/abs/2006SoPh..237..267M}
}

@article{arramy2026,
    author = {{Arramy}, Dimitri and {de la Cruz Rodr{\'\i}guez}, Jaime and {Pastor Yabar}, Adur and {Leenaarts}, Jorrit},
    title = "{On the inferred spatio-temporal evolution of magnetic fields during an X1-class flare}",
    journal = {\aap},
    year = 2026,
    status = {submitted, not accepted yet},
    
}

@ARTICLE{2022A&A...664A...8M,
       author = {{Morosin}, R. and {de la Cruz Rodr{\'\i}guez}, J. and {D{\'\i}az Baso}, C.~J. and {Leenaarts}, J.},
        title = "{Spatio-temporal analysis of chromospheric heating in a plage region}",
      journal = {\aap},
         year = 2022,
        month = aug,
       volume = {664},
          eid = {A8},
        pages = {A8},
          doi = {10.1051/0004-6361/202243461},
archivePrefix = {arXiv},
       eprint = {2203.01688},
 primaryClass = {astro-ph.SR},
       adsurl = {https://ui.adsabs.harvard.edu/abs/2022A&A...664A...8M}
}

@ARTICLE{Scharmer2026,
       author = {{Scharmer}, G.~B. and {de la Cruz Rodr{\'\i}guez}, J. and {Leenaarts}, J. and {Lindberg}, B. and {S{\"u}tterlin}, P. and {Hillberg}, T. and {Pietraszewski}, C. and {de Wijn}, A.~G. and {Foster}, M. and {Storey}, J.},
        title = "{The making of robust and highly performing imaging spectropolarimeters for large solar telescopes}",
      journal = {\aap},
         year = 2026,
        month = jan,
       volume = {705},
          eid = {A55},
        pages = {A55},
          doi = {10.1051/0004-6361/202555819},
archivePrefix = {arXiv},
       eprint = {2506.05143},
 primaryClass = {astro-ph.IM},
       adsurl = {https://ui.adsabs.harvard.edu/abs/2026A&A...705A..55S}
}

\begin{appendix}

\section{Electric current spatial and cumulative distributions at additional optical depths}\label{sec:AP_EC}

\subsection{Accuracy of the solver}

\begin{figure}[ht!]
    \centering
    \includegraphics[width=0.95\linewidth]{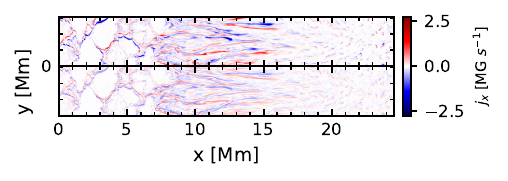}
    \caption{Map of solutions of $j_x$, the upper case is the MHD simulation, the lower is the solution from the solver , at $\mathrm{log}\left(\tau_{500}\right)=-1$.}
    \label{fig:AP_main_mapjx-1}
\end{figure}

\begin{figure}[ht!]
    \centering
    \includegraphics[width=0.95\linewidth]{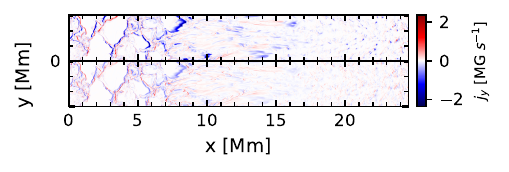}
    \caption{Same as Fig. \ref{fig:AP_main_mapjx-1} but for $j_y$.}
    \label{fig:AP_main_mapjy-1}
\end{figure}

\begin{figure}[ht!]
    \centering
    \includegraphics[width=0.95\linewidth]{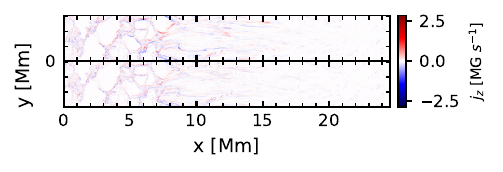}
    \caption{Same as Fig. \ref{fig:AP_main_mapjx-1} but for $j_z$.}
    \label{fig:AP_main_mapjz-1}
\end{figure}

\begin{figure}[ht!]
    \centering
    \includegraphics[width=0.95\linewidth]{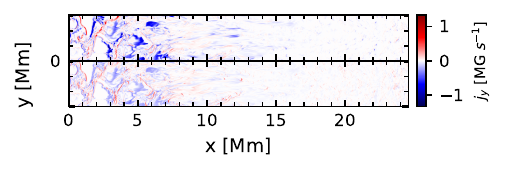}
    \caption{Map of solutions of $j_y$, the upper case is the MHD simulation, the lower is the solution from the solver , at $\mathrm{log}\left(\tau_{500}\right)=-2$.}
    \label{fig:AP_main_mapjy-2}
\end{figure}

\begin{figure}[ht!]
    \centering
    \includegraphics[width=0.95\linewidth]{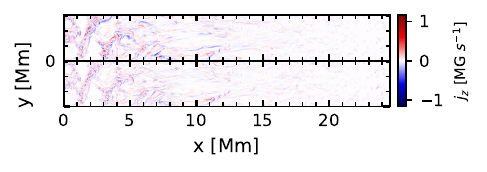}
    \caption{Same as Fig. \ref{fig:AP_main_mapjy-2} but for $j_z$.}
    \label{fig:AP_main_mapjz-2}
\end{figure}

\begin{figure}[ht!]
    \centering
    \includegraphics[width=0.95\linewidth]{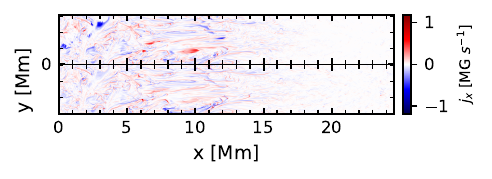}
    \caption{Map of solutions of $j_x$, the upper case is the MHD simulation, the lower is the solution from the solver , at $\mathrm{log}\left(\tau_{500}\right)=-3$.}
    \label{fig:AP_main_mapjx-3}
\end{figure}

\begin{figure}[ht!]
    \centering
    \includegraphics[width=0.95\linewidth]{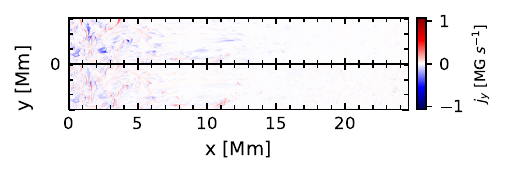}
    \caption{Same as Fig. \ref{fig:AP_main_mapjx-3} but for $j_y$.}
    \label{fig:AP_main_mapjy-3}
\end{figure}

\begin{figure}[ht!]
    \centering
    \includegraphics[width=0.95\linewidth]{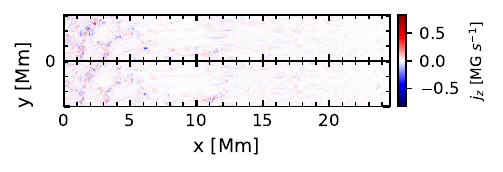}
    \caption{Same as Fig. \ref{fig:AP_main_mapjx-3} but for $j_z$.}
    \label{fig:AP_main_mapjz-3}
\end{figure}

Figures~\ref{fig:AP_main_mapjx-1} to~\ref{fig:AP_main_mapjz-3} show maps of $j_x$, $j_y$, and $j_z$ for the comparison of the MHD and HE3D cases at $\mathrm{log}\left(\tau_{500}\right)=-1,-2,-3$. These are complementary to the results described in Sect. \ref{subsec:maintest}.

\begin{figure}
\centering
\includegraphics[width=\linewidth]{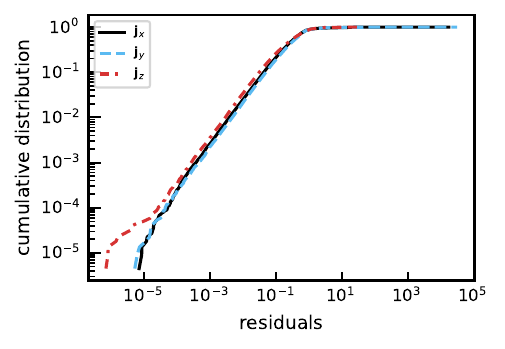}
\caption{Cumulative distribution functions for the three different components, with the relative differences between HE3D and MHD at $\mathrm{log}(\tau_{500})=-1$.}
\label{fig:AP_main_cdf_jx_all-1}
\end{figure}

\begin{figure}
\centering
\includegraphics[width=\linewidth]{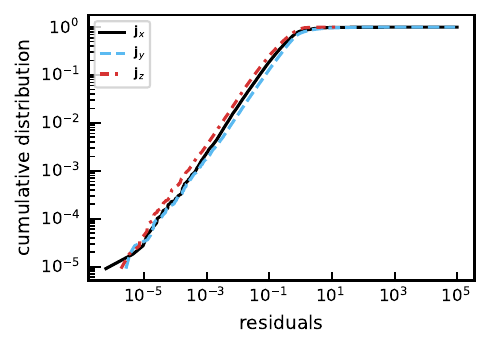}
\caption{Cumulative distribution functions for the three different components, with the relative differences between HE3D and MHD at $\mathrm{log}(\tau_{500})=-3$.}
\label{fig:AP_main_cdf_jx_all-3}
\end{figure}

Figures~\ref{fig:AP_main_cdf_jx_all-1} and~\ref{fig:AP_main_cdf_jx_all-3} show the cumulative distributions of relative differences for all three components at $\mathrm{log}(\tau_{500})=-1$ and $\mathrm{log}(\tau_{500})=-3$, respectively. The median relative differences for $\mathrm{log}\left(\tau_{500}\right)=-1$ are  $0.28$ (x), $0.30$ (y), $0.28$ (z) and $0.27$ (x), $0.37$ (y), $0.27$ (z) for $\mathrm{log}\left(\tau_{500}\right)=-3$.

\FloatBarrier 
\raggedbottom

\section{Heights of optical depth layers}\label{sec:AP_height}

\begin{figure}[H]
    \centering
    \includegraphics[width=\linewidth]{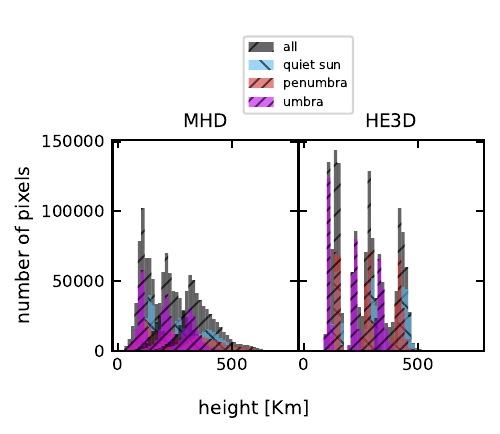}
    \caption{Distribution of height differences $\Delta^{i}_{z}$ with $i=[-1,-2,-3]$ for cases of MHD and HE3D, divided into quiet Sun, penumbra and umbra}
    \label{fig:AP_main_hist_z_reg}
\end{figure}

Figure~\ref{fig:AP_main_hist_z_reg} shows the distribution of distances between optical depth layers $\Delta_{z}^{i} = |z\left[\log_{10}\left(\tau_{500}\right)=i\right] - z\left[\log_{10}\left(\tau_{500}\right)=0\right]|$ similar to Fig.~\ref{fig:main_hist_z} with $i=[-1,-2,-3]$, but now divided into quiet Sun, penumbra and umbra to highlight the differences between these regions.

\FloatBarrier

\section{Spatial distribution of heights of optical depth layers}\label{sec:AP_heightdiff}

\begin{figure}[H]
    \centering
    \includegraphics[width=\linewidth]{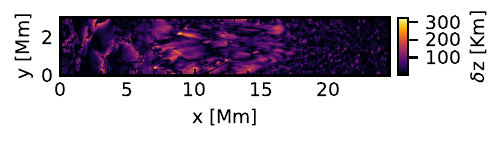}
    \caption{Map of change in height differences $\delta z=|\Delta_{z_{\mathrm{HE3D}}}^{-2}-\Delta_{z_{\mathrm{MHD}}}^{-2}|$ for the HE3D and MHD cases.}
    \label{fig:AP_main_heightdiff_map}
\end{figure}

Figure~\ref{fig:AP_main_heightdiff_map} on the other hand shows the spatial distribution of $\delta z=|\Delta_{z_{\mathrm{HE3D}}}^{-2}-\Delta_{z_{\mathrm{MHD}}}^{-2}|$. The largest deviations are found in the intergranular lanes and penumbra, which is expected due to the fact that terms neglected in Eq.~\ref{eq:HE} will play a strong role in these parts.

\FloatBarrier

\section{Wilson depression}

\begin{figure}[H]
    \centering
    \includegraphics[width=\linewidth]{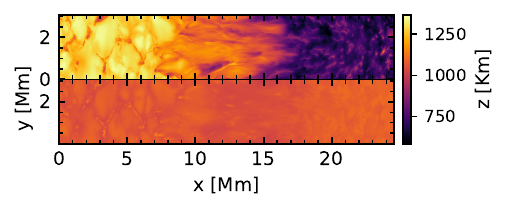}
\caption{Maps of height in kilometers at  $\mathrm{log}_{10}\left(\tau_{500}\right)=0$, i.e. Wilson depression for the MHD (upper) and HE3D (lower) case.}
    \label{fig:AP_main_wd0}
\end{figure}

Figure~\ref{fig:AP_main_wd0} shows the Wilson depression, which is a map of the absolute height within the simulation domain at an optical depth of $\mathrm{log}_{10}\left(\tau_{500}\right)=0$, for both MHD and HE3D. As discussed in Sect.~\ref{sec:4}, the HE3D case fails to recover the global structure that is present in the simulation, though the granules in the quiet sun are visible.  

\FloatBarrier 
\clearpage

\end{appendix}
\end{document}